\documentclass[reprint,
superscriptaddress,
amsmath,amssymb,
apl,
]{revtex4-2}
\usepackage[utf8]{inputenc}

\usepackage[usenames,dvipsnames]{color}
\usepackage{graphicx}
\graphicspath{{./images/} }
\usepackage{textcomp}
\usepackage{dcolumn}
\usepackage{bm}
\usepackage{comment}         
\usepackage[usenames,dvipsnames]{color}
\usepackage{url}

\usepackage{siunitx}     
\usepackage{amsmath}
\usepackage{tabularx}
\usepackage[table,xcdraw]{xcolor}
\usepackage{makecell}    
\usepackage{braket}      
\usepackage{bbold}       

\begin{document}
\title{Dual-rail encoding with superconducting cavities}

\author{James D. Teoh}
\email{james.teoh@yale.edu}
\affiliation{Departments of Applied Physics and Physics, Yale University, New Haven, CT, USA}
\affiliation{Yale Quantum Institute, Yale University, New Haven, CT, USA}

\author{Patrick Winkel}%
\email{patrick.winkel@yale.edu}
\affiliation{Departments of Applied Physics and Physics, Yale University, New Haven, CT, USA}
\affiliation{Yale Quantum Institute, Yale University, New Haven, CT, USA}

\author{Harshvardhan K. Babla}%
\affiliation{Departments of Applied Physics and Physics, Yale University, New Haven, CT, USA}
\affiliation{Yale Quantum Institute, Yale University, New Haven, CT, USA}

\author{Benjamin J. Chapman}%
\affiliation{Departments of Applied Physics and Physics, Yale University, New Haven, CT, USA}
\affiliation{Yale Quantum Institute, Yale University, New Haven, CT, USA}

\author{Jahan Claes}%
\affiliation{Departments of Applied Physics and Physics, Yale University, New Haven, CT, USA}
\affiliation{Yale Quantum Institute, Yale University, New Haven, CT, USA}

\author{Stijn J. de Graaf}%
\affiliation{Departments of Applied Physics and Physics, Yale University, New Haven, CT, USA}
\affiliation{Yale Quantum Institute, Yale University, New Haven, CT, USA}

\author{John W. O. Garmon}%
\affiliation{Departments of Applied Physics and Physics, Yale University, New Haven, CT, USA}
\affiliation{Yale Quantum Institute, Yale University, New Haven, CT, USA}

\author{William D. Kalfus}%
\affiliation{Departments of Applied Physics and Physics, Yale University, New Haven, CT, USA}
\affiliation{Yale Quantum Institute, Yale University, New Haven, CT, USA}

\author{Yao Lu}%
\affiliation{Departments of Applied Physics and Physics, Yale University, New Haven, CT, USA}
\affiliation{Yale Quantum Institute, Yale University, New Haven, CT, USA}

\author{Aniket Maiti}%
\affiliation{Departments of Applied Physics and Physics, Yale University, New Haven, CT, USA}
\affiliation{Yale Quantum Institute, Yale University, New Haven, CT, USA}

\author{Kaavya Sahay}%
\affiliation{Departments of Applied Physics and Physics, Yale University, New Haven, CT, USA}
\affiliation{Yale Quantum Institute, Yale University, New Haven, CT, USA}

\author{Neel Thakur}%
\affiliation{Departments of Applied Physics and Physics, Yale University, New Haven, CT, USA}
\affiliation{Yale Quantum Institute, Yale University, New Haven, CT, USA}

\author{Takahiro Tsunoda}%
\affiliation{Departments of Applied Physics and Physics, Yale University, New Haven, CT, USA}
\affiliation{Yale Quantum Institute, Yale University, New Haven, CT, USA}

\author{Sophia H. Xue}%
\affiliation{Departments of Applied Physics and Physics, Yale University, New Haven, CT, USA}
\affiliation{Yale Quantum Institute, Yale University, New Haven, CT, USA}

\author{Luigi Frunzio}%
\affiliation{Departments of Applied Physics and Physics, Yale University, New Haven, CT, USA}
\affiliation{Yale Quantum Institute, Yale University, New Haven, CT, USA}

\author{Steven M. Girvin}%
\affiliation{Departments of Applied Physics and Physics, Yale University, New Haven, CT, USA}
\affiliation{Yale Quantum Institute, Yale University, New Haven, CT, USA}

\author{Shruti Puri}%
\affiliation{Departments of Applied Physics and Physics, Yale University, New Haven, CT, USA}
\affiliation{Yale Quantum Institute, Yale University, New Haven, CT, USA}

\author{Robert J. Schoelkopf}%
\email{robert.schoelkopf@yale.edu}
\affiliation{Departments of Applied Physics and Physics, Yale University, New Haven, CT, USA}
\affiliation{Yale Quantum Institute, Yale University, New Haven, CT, USA}

\date{\today}

\begin{abstract}
The design of quantum hardware that reduces and mitigates errors is essential for practical quantum error correction (QEC) and useful quantum computation. To this end, we introduce the circuit-Quantum Electrodynamics (QED) dual-rail qubit in which our physical qubit is encoded in the single-photon subspace, $\{\ket{01}, \ket{10}\}$, of two superconducting microwave cavities. The dominant photon loss errors can be detected and converted into erasure errors, which are in general much easier to correct. In contrast to linear optics, a circuit-QED implementation of the dual-rail code offers unique capabilities. Using just one additional transmon ancilla per dual-rail qubit, we describe how to perform a gate-based set of universal operations that includes state preparation, logical readout, and parametrizable single and two-qubit gates. Moreover, first-order hardware errors in the cavities and the transmon can be detected and converted to erasure errors in all operations, leaving background Pauli errors that are orders of magnitude smaller. Hence, the dual-rail cavity qubit exhibits a favorable hierarchy of error rates and is expected to perform well below the relevant QEC thresholds with today’s coherence times.
\end{abstract}
\maketitle

\section{Introduction}
There has been remarkable progress in the physical implementation of quantum information processing devices over the past two decades. Several platforms, including trapped ions\,\cite{Cirac_1995,Monz_2011,Leibfried_2003,Blatt_2012,Haffner_2008}, neutral atoms\,\cite{Bloch_2008,Saffman_2010,Bloch_2012,Saffman_2016}, and superconducting circuits\,\cite{Wallraff_2004,BlaisCQED}, have advanced to the stage where systems with dozens or hundreds of physical qubits can be assembled and programmed \cite{Arute_2019,Xu_2022_MQP} to perform simple algorithms. But even for the so-called Noisy Intermediate Scale Quantum (NISQ) applications\,\cite{preskill_2018} that are being investigated today, significant improvements are necessary in the error rates for all types of operations, including initialization and measurement as well as single and two-qubit quantum gates. Moving beyond the current NISQ era will require the implementation of quantum error correction that performs well enough to realize significant gains in logical fidelity. 
While some of the performance levels of physical qubits are now approaching the theoretical thresholds that are required\,\cite{Google_EC_gain_2022}, when and how practical error correction might be achieved remains an outstanding question.
\\
\\
Quantum error correction (QEC) is challenging because of many simultaneous requirements.
These include a significant overhead in the number of physical qubits used to encode logical information, the rapid operation of a complex sequence of gates and measurements to detect errors, and the high fidelity and precision of the components and operations that make up the system. Finding more efficient schemes for QEC that can ease these requirements is a very active area of current research. A wide range of approaches has been explored, including more efficient codes\,\cite{LDPC_code_1962}, qubits that have structure or bias\,\cite{Mirrahimi_2014,Puri_2019,Grimm_2020,Jaya_2022, Frattini_2022,Chavez_2022, Lukin2022_SP} in their noise processes and the modifications of encoding schemes to utilize this structure\,\cite{Aliferis_2008,BonillaAtaides_2021,Darmawan_2021,Chamberland_2022,Guillaud_2019_Rep,AWS_transmonDR2022,Roffe2022_LDPCnoisebias,Shim2016,OliverPRX2020}, code switching or ``pieceable" constructions\,\cite{YoderPieceable2016,Lin_Pieceable_2021} that do not require fully transversal operations, and correction schemes that can flag\,\cite{Chao_2018,Postler2022} certain errors. One can even build redundancy into a single physical system, replacing the usual qubits with multi-level systems such as qutrits\,\cite{Gokhale_2019}. Bosonic encodings\,\cite{Weizhou_2021} such as the GKP code\,\cite{GKPOG,Fluhmann2019,Campagne-Ibarcq2020,VladGKP2022,Grimson_2021_GKP}, cat codes\,\cite{haroche2006exploring,Vlastakis_2013,Mirrahimi_2014,Legthas_2015,Ofek2016}, or binomial codes\,\cite{MichaelBin2016,SunBin2019,Xu_2020_CPhase,Ni_2022_binomial} have proven to be especially hardware efficient.
\\
\\
When choosing an error correction architecture, it is important to remember that there is a natural hierarchy of physical and logical errors. Not all errors occur with equal probability nor are they equally harmful to the code. On the logical level, detected erasure errors are especially benign, having both significantly higher thresholds and more favorable scaling with code distance\,\cite{GrasslErasure1997,Gottesman1997_SP,StaceSurface50Erasure2009,StaceSurface25Erasure2009,Delfosse2020}. Pauli errors, the most well-studied error model, are also correctable but have lower thresholds and a less favorable scaling with code distance. Lastly, the accumulation of undetected leakage errors out of the code space is an uncorrectable error and therefore most damaging\,\cite{Aliferis_2009,Fowler_2013_leakage,Ghosh_2015_leakage,Suchara_2015,Bultink_2020_leakage,Varbanov2020,McEwen2021,Krinner2022}. Aligning the physical behavior of gates and qubits to respect this hierarchy is therefore another important strategy for efficient QEC. 
Erasures are the dominant errors for some platforms such as linear quantum optics\,\cite{Li_2015_FTLO,Rudolph2017,FusionBased2021}. An important recent insight\,\cite{wu_2022} is that one may also choose or design physical qubits and operations where additional measurements can detect certain errors and then allow them to be converted to erasures, easing the thresholds and overhead requirements for QEC.
\\
\\
Superconducting microwave cavities are attractive candidates for hardware-efficient error correction. Microwave resonators can be implemented in planar or 3-D geometries, and typically have longer lifetimes\,\cite{Axline2016coaxline,Reagor2016,Romanenko_2017,milul2023superconducting} than nonlinear qubits containing Josephson junctions. Moreover, they exhibit little or no intrinsic dephasing\,\cite{Rosenblum2018,ECD2022,VladGKP2022,milul2023superconducting}, with the predominant error mechanism simply being photon loss. This means they naturally have a type of bias in their decoherence. Over the last few years, these properties have allowed experiments that demonstrate quantum memories\,\cite{Ofek2016,Weizhou_2021,VladGKP2022}, single qubit quantum gates\,\cite{OCTHeeres2017, Reinhold2020}, and remote entanglement\,\cite{Burkhart2020} where the error correction approaches or even exceeds the breakeven point and logical fidelity begins to improve. Just as qubit based codes incur an overhead because they require multiple physical qubits per logical qubit and additional measurement ancillae, bosonic codes do incur an increase in the decay rate proportional to the number of photons stored in the system. In addition, transmons or other traditional superconducting qubits are still required to enable the nonlinear operations necessary for control and measurement in the codespace. These ancillae then introduce additional error channels that can limit fidelity unless they can be managed and prevented from propagating to the bosonic logical qubit\,\cite{Rosenblum2018,Puri_2019,Reinhold2020}.
\\
\\
Here we propose a platform for bosonic error correction by combining the previously developed techniques of circuit QED and an implementation of the dual-rail encoding with a single photon stored in a pair of coupled microwave cavities. Employing microwave cavities for the dual-rail encoding uses a small number of photons and preserves the noise bias, where photon loss only appears as a detectable erasure. We argue that this system can have a remarkable and highly favorable hierarchy of error rates, with Pauli errors that are orders of magnitude weaker than erasures, and leakage that is even smaller. In addition, we describe a complete set of one and two-qubit operations that use a beamsplitter interaction between the cavities and a dispersively coupled transmon ancilla to enable high-fidelity preparation, measurement, and non-linear control. We show how the dominant physical errors, including decay, heating, and dephasing in the transmon ancilla, can be effectively detected and converted to erasures, with only second-order contributions to a residual rate of Pauli errors. Even with current levels of decoherence and previously demonstrated operation speeds\,\cite{YaoBS2022, StijnBS2022}, this scheme will allow post-selected algorithms with fidelity significantly beyond that of today's NISQ machines. Finally, this approach should perform well in concatenated QEC schemes such as the surface code\,\cite{Kitaev_1998,Dennis_2002,Fowler_2012,BonillaAtaides_2021,wu_2022}, enabling a faster path to fault-tolerant quantum computing. 

\section{Results}
\subsection{Dual-rail cavity qubit}
\begin{figure}[t]
	\includegraphics[width=0.9\linewidth]{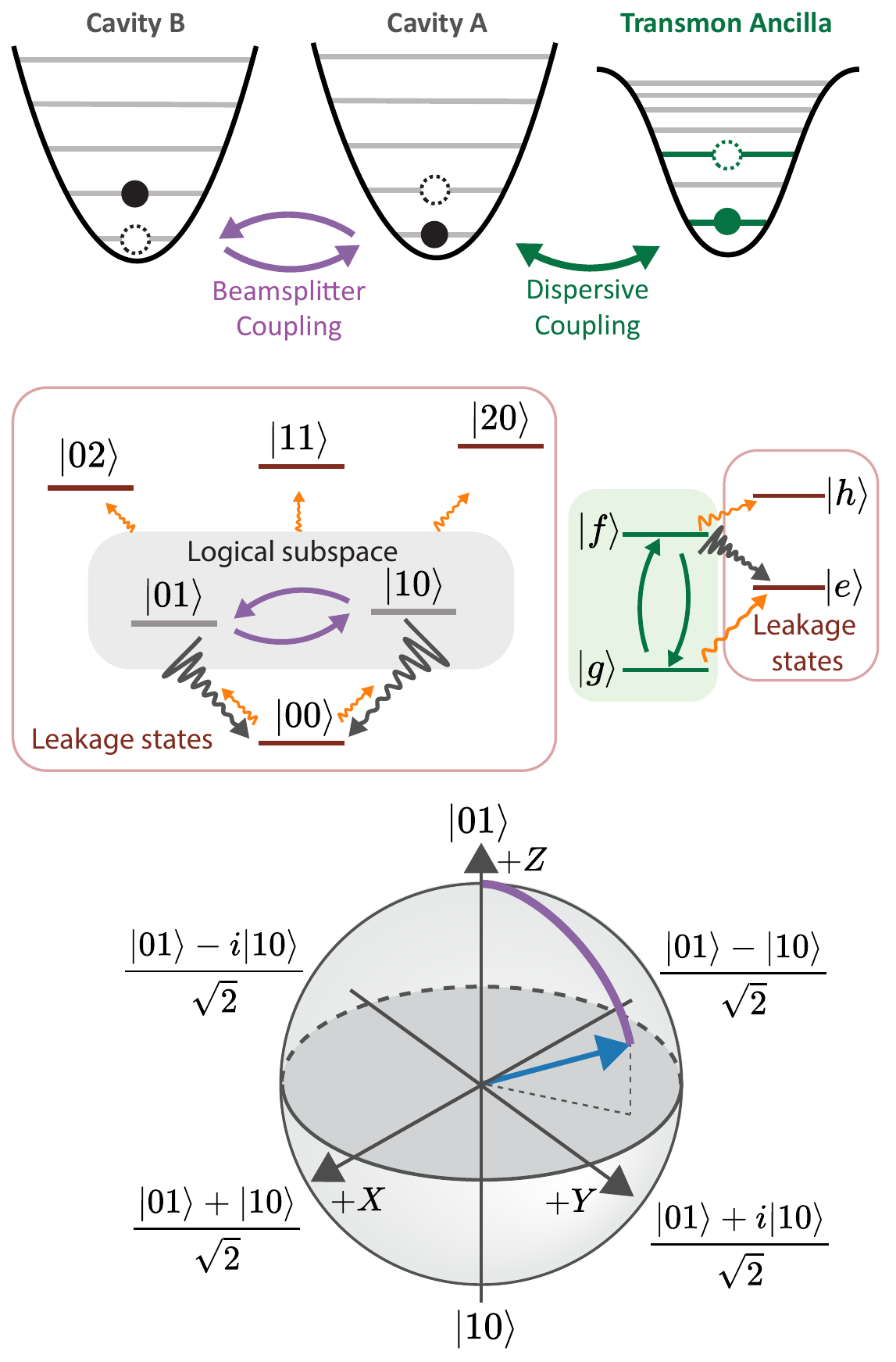}
	\caption{ 
        The cavity dual-rail qubit is composed of two standing superconducting cavities taking the role of the two waveguide modes in linear quantum optics. The cavities are coupled by a switchable beamsplitter interaction. The logical code words are encoded in the single-photon subspace of the cavity system, with $\ket{0}_{\mathrm{L}} = \ket{01}$ and $\ket{1}_{\mathrm{L}} = \ket{10}$, for which the beamsplitter interaction is sufficient to facilitate arbitrary single-qubit rotations. Energy relaxation events into the common ground state $\ket{00}$, as well as single heating events into the higher energy states $\ket{11}$, $\ket{20}$ and $\ket{02}$ bring the system into a leakage space distinguished from the logical subspace by the joint photon-number parity. Therefore, the dominant error in the cavity system, single-photon decay, can be converted into erasure errors by measuring the joint-parity. A single transmon ancilla is dispersively coupled to one of the cavities and acts as a resource for non-Gaussian operations. By operating the transmon in the g-f-manifold, we are able to detect decay events in the ancilla, expected to be the most dominant error in the system for state-of-the art realizations, and convert them into erasure errors as well.     
        }
		\label{fig:dual_rail_encoding}
\end{figure}

The concept of the superconducting dual-rail cavity qubit is based on the intrinsic noise bias of superconducting cavities, for which single-photon loss is the dominant physical error channel and the pure dephasing caused by fluctuations in cavity frequency is orders of magnitude smaller, at least for an isolated cavity. Furthermore, the single-photon loss rates for superconducting cavities are among the smallest error rates measured in any superconducting device\,\cite{Reagor_2013} and are likely to improve at least as rapidly as qubits based on Josephson junctions with continuing advances in material science and fabrication.
\\
\\
The dual-rail superconducting qubit we present here has a similar encoding to that used in linear quantum optics platforms\,\cite{Chuang_1995}, where logical information is encoded in two distinguishable bosonic modes, but with access to a set of quantum non-demolition (QND) measurements and nonlinear controls afforded by the circuit quantum electrodynamics (cQED)\,\cite{BlaisCQED, Koch_2007} toolkit. Crucially, we can use the non-linearity of a transmon (or other) ancilla to directly perform two-qubit entangling gates, and also QND measurements such as joint photon-number parity 
measurements on two cavities. These capabilities mean, in contrast to other implementations, we can realize a gate-based approach for the dual-rail qubit with no need for heralding, with the ability to detect many types of physical errors at the hardware level.
\\
\\
Dual-rail encodings based on superconducting qubits have been previously suggested\,\cite{Shim2016,AWS_transmonDR2022} and implemented using transmon qubits\,\cite{OliverPRX2020,levine2023demonstrating}. Unlike microwave cavities, transmons have little inherent noise-bias, with dephasing rates often comparable to their decay rates. While we could detect decay events in this transmon dual-rail qubit, dephasing will remain the dominant error source, introducing both Pauli X and Z errors. 
In recent implementations\,\cite{OliverPRX2020,levine2023demonstrating} the desired noise-bias was engineered by tuning two transmons on resonance with each other. 
\\
\\
Instead of two propagating optical modes, we choose the two rails to be superconducting standing-wave cavity modes with distinct microwave frequencies. These could be two modes of the same resonator\,\cite{ZakkaBajjani2011}, but here we will discuss the case where they are in separate but adjacent resonators, as shown in Fig.\,\ref{fig:dual_rail_encoding}.\newpage
\vfill

The logical codewords are $\ket{0}_{\mathrm{L}} = \ket{01}$ and $\ket{1}_{\mathrm{L}} = \ket{10}$.  By encoding the logical state of the qubit in the single-photon subspace of the joint cavity Hilbert space with odd joint photon-number parity, we are able to convert the dominant errors in the cavity system, namely single-photon loss to the common ground state $\ket{00}$, into detectable erasure errors by means of joint-parity measurements. In addition, photon gain events in the cavities, which are generally rare in thermal equilibrium, are also detectable by the same measurement, leaving cavity dephasing as the dominant Pauli error in the system. The error probability and scaling during idling time is summarized in Table\,\ref{Tab_errors_idling_time}. This shows a remarkable physical error hierarchy that follows the optimal structure for quantum error correction codes that include erasure conversion. In following sections, we will see that Pauli errors remain small even when we introduce the transmon ancilla and cavity-cavity coupler needed for a full set of operations. \\
\\
Compared to other bosonic codes\,\cite{GKPOG,LeghtasCatCode, MichaelBin2016}, there is little increase in photon number when using the dual-rail encoding relative to the simple ‘single-rail’, or Fock $\{\ket{0},\ket{1}\}$ encoding. The main cost is we now need two cavity modes per qubit instead of one, and we double the number of photons (from $\bar{n}=0.5$ to $\bar{n} = 1$). Single qubit gates for Fock $\{\ket{0},\ket{1}\}$ cavity qubits are more difficult and slower, requiring extensive use of a transmon ancilla\,\cite{OCTHeeres2017}. In a dual-rail cavity qubit, arbitrary single-qubit rotations are achieved with a simple switchable beamsplitter interaction between the cavities, by pumping a dedicated non-linear coupler\,\cite{StijnBS2022,YaoBS2022}. This is another vast simplification compared to other bosonic encodings. \\
\\
Since the (frequency-converting) beamsplitter interaction is activated using parametric processes, the cavities are not on resonance during the idling time, resulting in a large on-off ratio. The amplitude and phase of the beamsplitter pumps are analogous to the usual RF drives used to control transmon qubits, enabling techniques such as dynamical decoupling\,\cite{Bylander2011,Yan2013} to further suppress dephasing errors, as well as unwanted effects such as no-jump backaction arising from differences in cavity single-photon decay rates (see App.\,\ref{App:no_jump}). \\
\\
Non-Gaussian operations on the dual-rail cavity qubit use the transmon ancilla as the source of non-linearity. This includes state preparation, logical readout, QND cavity loss (joint-parity) detection and the two-qubit entangling gates. Since transmon errors occur more frequently than all cavity idling errors, we design protocols to preserve the error hierarchy that exists for idle dual-rail qubits by detecting first-order transmon errors after the sequence.

\begin{table*}[t]
\caption{The dual-rail cavity qubit is composed of multiple hardware components - two superconducting cavities, a single parametric coupler, and a single transmon ancilla, as shown in Fig.\,\ref{fig:dual_rail_encoding}. To estimate the physical error rate of the proposed dual-rail qubit during idling time, and the logical errors in the encoded qubit they can cause, we assume typical performances achievable for the individual components today. 
By construction, the hierarchy of these events, i.e. their probability of occurrence, aligns with the desired scaling for an outer QEC layer: The dominant error is photon loss out of the computational space caused by cavity relaxation at rate $\bar{\kappa} = (\kappa_{\mathrm{a}} + \kappa_{\mathrm{b}}) / 2$, which, by measuring the joint-parity (JP), can be converted into erasure errors. As a consequence, erasure errors, though still rare, are significantly more likely than Pauli errors. During the idling time, phase flip events are predominantly caused by dephasing events in the cavity system or heating events in the transmon, where $\bar{n}_{\mathrm{th,A}}$ is the average thermal ancilla population and $\Gamma_{1,\mathrm{A}}$ the energy relaxation rate.
By measuring the transmon (M) and ensuring it remains in its ground state we can detect ancilla heating events, at least to first order.
No-jump backaction arises when there is a finite difference between the single photon decay rates of the cavities given by $\Delta \kappa = \kappa_{\mathrm{b}} - \kappa_{\mathrm{a}}$. Over time, the weighting of a dual-rail qubit superposition changes to increase the probability of the photon being in the longer lived cavity, despite no photon loss being detected. This is a second-order effect, and is a type of phase-flip type error. Unlike cavity dephasing, this error can change the qubit's populations.
The closest error that resembles a bit flip error is the unlikely combination of a cavity decay and heating event, resulting in an indirect transition between the logical codewords at extremely low rates. Leakage events undetectable by the joint-parity measurement are extremely rare, as they require two heating events in the cavities to bring the dual-rail qubit into a state with odd joint-parity outside the logical codespace. However, these events could still be detected with a QND measurement\cite{TakaGates2022} of the joint ``super-parity," (JSP) as they have a different value for $(n_a +n_b) \text{mod} 4$. For the calculation of the error probabilities, we assume single-photon decay rates of $\kappa_{\mathrm{a}} = (1.5\,\mathrm{ms})^{-1}$ and $\kappa_{\mathrm{b}} = (0.5\,\mathrm{ms})^{-1}$ for the cavities,  $\Gamma_{1,\mathrm{A}} = (100\,\si{\micro\second})^{-1}$ for the ancilla transmon, pure dephasing rates $\gamma_{\varphi,\mathrm{a}} = \gamma_{\varphi,\mathrm{b}} = (20\,\mathrm{ms})^{-1}$, and a thermal population $\bar{n}_{\mathrm{th}} = \bar{n}_{\mathrm{th,A}} = 0.01$ for all quantum elements. The noise bias of the dual-rail specifically refers to erasure errors happening much more frequently than all other types of errors such as Pauli and leakage errors. The noise bias is quantified for each error process by comparing to the cavity photon loss error rate, the dominant source of erasure errors in the system when idling.}
\label{Tab_errors_idling_time}
\small
\centering
\renewcommand{\arraystretch}{1.5}
\begin{tabular}{l||c|c|c|c|c|c}
\toprule
Error process & Scaling & Probability in $1 \, \si{\micro\second}$ & Noise bias & Effective lifetime & Error type & Detection \\ \hline \hline
cavity photon loss & $\Bar{\kappa} t$ & $10^{-3}$ & 1 & $1\,\si{\milli\second}$ & Erasure & JP\\
cavity heating &  $\bar{n}_{\mathrm{th}}\Bar{\kappa} t$ & $10^{-5}$ & $10^{2}$ & $100\,\si{\milli\second}$ & Erasure & JP\\
cavity dephasing & $\gamma_{\varphi} t$ & $10^{-4}$ & $10^{1}$ & $10\,\si{\milli\second}$ & Phase flip & - \\
ancilla heating &  $\bar{n}_{\mathrm{th,A}} \Gamma_{1,\mathrm{A}} t$ & $10^{-4}$ & $10^{1}$ & $10\,\si{\milli\second}$ & Phase flip & M \\
no-jump backaction &  $\left(\frac{1}{4} \Delta \kappa t \right)^2$ & $10^{-6}$ & $10^{3}$ & $1\,\si{\second}$ & Phase flip & - \\
cavity photon loss + heating  &  $ \bar{n}_{\mathrm{th}} \left(\bar{\kappa} t \right)^2$& $10^{-8}$ & $10^{5}$ & $100\,\si{\second}$ & Bit flip & - \\
cavity heating $\times 2$  &  $3 \left(\bar{n}_{\mathrm{th}} \bar{\kappa} t \right)^2$ & $10^{-9}$ & $10^{6}$ & $\sim 1\,\si{\hour}$ & Leakage & JSP
\end{tabular}
\end{table*}
\subsection{State preparation and measurement}
We now describe how to perform preparation and readout of logical states on a dual-rail qubit, which may also be constructed to avoid first-order sensitivity to most physical errors. The general approach is to use the dispersive interaction between one of the cavities and the ancilla transmon to learn logical state information from repeated photon-number parity measurements. We may then majority vote on many rounds of measurement outcomes to reach high assignment fidelity\,\cite{SalElder2020}  by using the most common outcome to decide the overall measurement result. This exponentially suppresses the effects of transmon errors and transmon readout errors at the cost of a small increase in the ongoing erasure rate due to cavity photon loss. For the proposed measurement scheme, logical readout fidelity for a dual-rail qubit is expected to be better than any other known qubit platform.\\
\\
\\
The QND nature of parity measurements performed on single cavities\,\cite{Sun_2014} means we can also use them to verify logical state preparation to very high fidelity. Preparing a state in the logical Z basis is simply a matter of loading a single photon into one of the two cavities. This may be done via optimal control pulses\,\cite{OCTHeeres2017} or cavity-transmon sideband drives\,\cite{Wallraff2007sidebands,Premaratne,Rosenblum2018CNOT,SalElder2020} and then verified with subsequent parity measurements to achieve error-robust state preparation with success probability $>99\%$. Unsuccessful state preparation may be treated as an erased qubit, or we can allow for multiple attempts at state preparation to boost the success probability. Due to low idling error rates of the dual-rail (Table.\,\ref{Tab_errors_idling_time}), the additional measurement time need not introduce logical errors at a substantial rate. Once again, the dominant errors are detectable leakage errors, i.e. erasures.
\\
\begin{figure*}[t]
	\includegraphics[width=1 \linewidth] {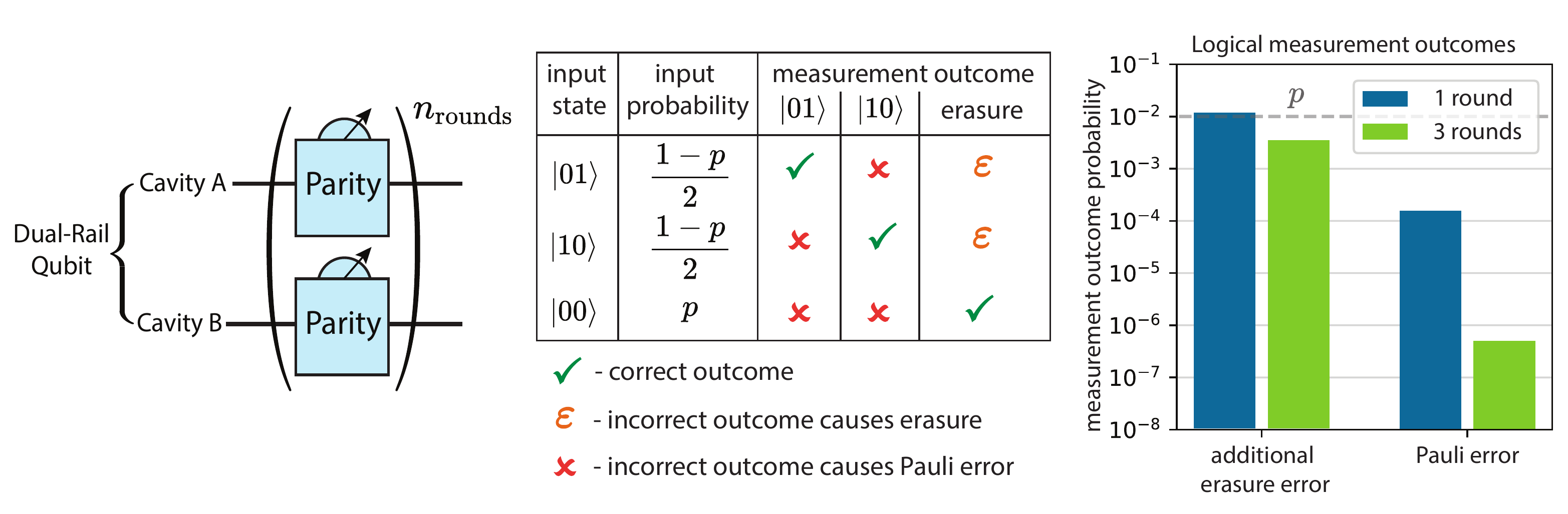}
	\caption{ 
The logical state of the dual-rail cavity qubit can be extracted from local parity measurements of the individual cavities. The table shows the three most likely input states $\ket{01}$, $\ket{10}$ and $\ket{00}$ together with their input probability, where $p$ is the probability of a photon loss event prior to the measurement, leaving the system in $\ket{00}$. We set $p=1\%$ which is the erasure probability due to photon loss in the cavities after approximately ten two-qubit gates. Depending on the input state, the outcome of the measurement is either correctly assigned to the actual input state (green ticks), incorrectly flagged as an erasure (orange), or incorrectly assigned to the logical states (red cross) resulting in a Pauli error. By repeating the measurement protocol $n$ times, the measurement errors can be significantly reduced, as shown in the right panel. For a single round of measurements (blue bars), ancilla readout errors are mainly converted into false positive erasure errors. With three rounds of measurements and majority voting (see App.\,\ref{sec:msmt_strat}), we become much more robust to transmon errors, leading to a factor of five reduction in the additional erasures and orders of magnitude reduction in the logical assignment error. As we increase the number of measurement rounds even further, the total measurement time gets longer, and so the probability of cavity photon loss increases, eventually increasing the chance of erasure again (not shown). Hence, due to finite transmon coherence times and readout fidelities, there can be an improvement in measurement performance when using 3 rounds instead of 1 round. The simulation models transmon decay during readout and other measurement errors (see App.\,\ref{sec:msmt_model}). We set  $T_1^{ge} = 100\,\si{\micro\second}$, $T_{\varphi}^{ee} = 100\,\si{\micro\second}$, and $\kappa_{\mathrm{a}} = \kappa_{\mathrm{b}} = (1\,\si{\milli\second})^{-1}.$}
		\label{fig:state_prep_measurement}
\end{figure*}
\\
For logical readout of a dual-rail qubit, we must ascertain whether a photon is in cavity Alice, $\ket{0}_{\mathrm{L}}$, or Bob, $\ket{1}_{\mathrm{L}}$, or in the case of a leakage event due to photon loss, neither cavity.  One simple way to do this is to measure the photon number parity in both cavities sequentially using the g-e-manifold of the transmon ancilla\,\cite{Vlastakis_2013,Sun_2014} which requires a time $\sim1/\chi_{\mathrm{ge}}\sim0.1-1\ \si{\micro\second}$ and is shown in Fig.\,\ref{fig:state_prep_measurement}. We may measure the parity sequentially using the same ancilla and swapping the cavity states in between each parity measurement, or an additional ancilla transmon for the second cavity can be added to the architecture if desired. If we are in the logical $\ket{0}_{\mathrm{L}}$ state we should obtain the transmon measurement string $(e_{\mathrm{a}},g_{\mathrm{b}})$ since we read out state $\ket{e}$ when there are an odd number of photons in the cavity. In the case of prior photon loss we obtain $(g_{\mathrm{a}},g_{\mathrm{b}})$ and flag the measurement as an erasure.\\
\\
The probability of mistaking $\ket{0}_{\mathrm{L}}$ for $\ket{1}_{\mathrm{L}}$ (and vice versa) should be exceedingly small even in the presence of decoherence. First-order errors in the transmon and cavities give measurement outcomes $(g_{\mathrm{a}},g_{\mathrm{b}})$ and $(e_{\mathrm{a}},e_{\mathrm{b}})$ which we flag as additional erasure errors. A combination of at least two errors is necessary for logical misassignment, a probability we estimate to be below $10^{-4}$ for realistic coherence times and transmon readout fidelities\,\cite{Evan2014_readout,WallraffReadout2017}. \\
\\
Dephasing on the cavity is not an issue when preparing and measuring dual-rail qubits in the Z basis, so we may operate the transmon ancilla in the g-e-manifold without the need to flag transmon errors.  Even though transmon decay can corrupt an individual parity measurement outcome, we may still correctly measure logical information via subsequent parity measurements to increase readout fidelity. Alternatively, we can map information about the cavity states onto the transmon using photon-number selective $\pi$ pulses. In combination with shelving the transmon excitation into higher states, we expect to achieve high logical readout fidelities even with a single round of measurements\,\cite{SalElder2020}.
\\
\\
By repeating each round of parity measurements three times, we can majority vote on the outcomes to further suppress the effects of transmon errors. Practically, this means fewer unnecessary erasure flags due to transmon errors along with a small increase in erasure flags due to cavity photon loss. We also note that majority voting suppresses the infidelity of the transmon measurements themselves. Here, we consider the consequence of two types of measurement errors: a misassignment of the transmon state due to the finite SNR of the measurement, and transmon decay during readout. In both cases, the inferred transmon state does not coincide with the physical state, which can corrupt the subsequent measurement outcome after resetting into the ground state\,\cite{Egger2018_reset}. Alternatively, we can use an unconditional reset\,\cite{Sunada2022_reset}, or no reset at all and instead update the expected measurement outcomes. A quantitative comparison of two particular measurement strategies is shown in Fig.\,\ref{fig:state_prep_measurement}, assuming state-of-the art experimental hardware.
\\
\\
In general, the optimal measurement strategy is highly context and device dependent, with tradeoffs between total measurement time, additional erasure rate and measurement fidelity. Measurement of ancilla dual-rail qubits in a surface code may prioritize measurement speed over assignment fidelity whereas post-selected short-depth circuits may favor assignment fidelity above all else. More measurement strategies are discussed in App.\,\ref{sec:msmt_strat}. The examples in Fig.\,\ref{fig:state_prep_measurement} show the minimum needed to suppress transmon errors and the benefits from majority voting. An experimental implementation of the state preparation and logical readout of a dual-rail cavity qubit has been recently reported in Ref.\,\cite{Chou2023_SPAM}, demonstrating a logical misassignment error of $(1.7\pm 0.3) \times 10^{-4}$ for a single round and $(3\pm1) \times 10^{-5}$ for two rounds of measurement.  

\begin{figure*}[]
	\includegraphics[width=0.8 \linewidth] {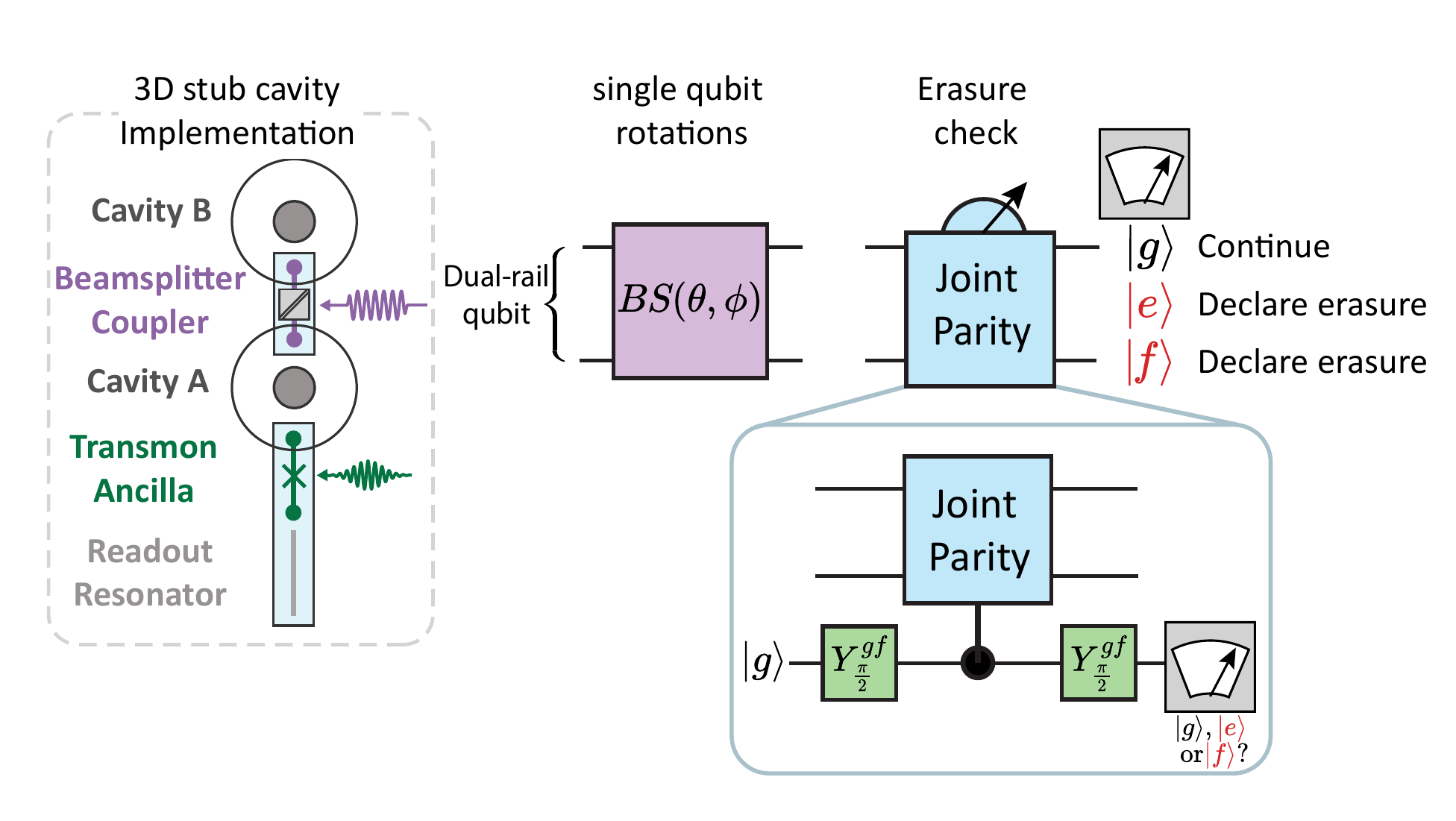}
	\caption{ 
        An example of hardware to implement one dual-rail cavity qubit (see Fig.\,\ref{fig:dual_rail_encoding}), shown here as 3D stub cavities, which have already demonstrated\,\cite{Reagor2016} single-photon decay times beyond $1\,\si{\milli\second}$, even in the presence of additional non-linear modes, and have intrinsically low pure dephasing rates reaching at least tens of milliseconds\,\cite{Rosenblum2018,ECD2022,VladGKP2022}. The beamsplitter interaction between the cavities, which is sufficient to implement arbitrary single-qubit rotations in the dual-rail subspace, can be realized with one of several choices of parametrically driven three-wave or four-wave mixing elements (violet), as recently demonstrated in Ref.\,\cite{YaoBS2022} and Ref.\,\cite{StijnBS2022}, respectively, without degrading the intrinsic coherence of the cavities. For the non-Gaussian operation, a transmon ancilla is capacitively coupled to one of the stub cavities and is operated in the g-f-manifold to allow for ancilla decay detection with a dedicated readout resonator. A key ingredient for the operation of the dual-rail qubit is the detection of physical errors in the quantum hardware, allowing us to convert leakage events into erasure errors. Such an erasure check is implemented by mapping the joint photon-number parity on the state of the ancilla. The ancilla starting in the ground state $\ket{g}$ is initialized in the $\ket{+X}$ state with a $\pi/2$ pulse. While photon-number parity measurements in a single cavity have been already demonstrated, we can measure the joint-parity by activating the beamsplitter interaction between the cavities at the same time. The second $\pi/2$ pulse maps the odd joint-parity states on the ground state $\ket{g}$ of the ancilla and the even states on $\ket{f}$. A decay of the ancilla during the operation is flagged by the final state $\ket{e}$. More details on the implementation of these composite operations are found in Ref.\,\cite{TakaGates2022}.            
}
		\label{fig:erasure_check}
\end{figure*}

\subsection{Single-qubit rotations}
\label{SEC:single_qubit}
Single-qubit gates in the dual-rail subspace are realized by a simple beamsplitter interaction between the cavities using a dedicated coupling element\,\cite{YaoBS2022,StijnBS2022}. The combination of beamsplitter strength and duration, which are controlled by microwave pump signals in an experiment, define the polar angle of the rotation, while arbitrary rotations around the Z-axis can be implemented in software by changing the phase of the pump(s). This parametrizable control is in many  ways analogous to typical use of RF signals for transmon single qubit gates. A SWAP gate between the cavities is equivalent to a $\pi$-pulse and a 50-50 beamsplitter corresponds to a $\pi/2$-pulse. Changing the pump phase changes the equator axis we rotate around on the Bloch sphere. Techniques for suppressing control errors, such as composite pulses, can be directly transferred to our beamsplitter control\,\cite{CompPulseChuang2004,NMRcomp2007}. Moreover, we can also make use of dynamical decoupling\,\cite{Bylander2011,Yan2013} to further enhance the dephasing time of our dual-rail qubit.  \\
\\
High speed beamsplitters\,\cite{StijnBS2022} and randomized benchmarking\,\cite{YaoBS2022} of single qubit gates in the dual-rail subspace have recently been experimentally demonstrated that approach 99.98\% gate fidelity when combined with an erasure check, in times of $\sim 50-100\,\mathrm{ns}$ per gate. This is on par with single qubit gates in transmons\,\cite{ChowGates2010}. Unlike a transmon qubit, there is no inherent speed limit set by the finite anharmonicity of the energy spectrum\,\cite{Motzoi_2009,Gambetta_DRAG_2011}, which increases leakage out of the computational space as we decrease gate times. It is also observed that the beamsplitter interaction does not induce any additional leakage errors out of the dual-rail subspace, aside from the expected (and detectable) decay to the $\ket{00}$ state set by the bare cavity single-photon decay rates. \\
\\
Using a parametric beamsplitter coupler allows our cavity modes to have large detunings, with vanishingly small swapping rates between the cavities in the absence of the pumps. Moreover, any static coupling between the cavities (within or between dual-rail qubits) simply leads to a renormalization of the cavity frequencies via mode hybridization, and is automatically compensated by calibrating to the dressed values that are observed in an experiment.\\
\\
The ease of arbitrary single-qubit rotations in the dual-rail encoding is a huge advantage compared to other error-correctable bosonic codes such as GKP\,\cite{GKPOG}, Binomial\,\cite{MichaelBin2016} or 4-legged cats\,\cite{haroche2006exploring, Mirrahimi_2014}. Whilst universal control of a harmonic oscillator via a transmon ancilla has been shown in many frameworks\,\cite{HeeresSNAP2015,OCTHeeres2017,ECD2022}, these approaches are comparatively slow to execute, require numerical optimization and, most importantly, are highly susceptible to ancilla transmon errors. \\
\\

\begin{table*}[t]
\centering
\caption{Physical error events in the quantum hardware during the single and two-qubit operations give rise to erasure and Pauli errors in the logical encoding. The probability of such events is determined by their rate of occurrence and the time interval considered, typically the two-qubit gate duration. For the cavity dual-rail qubit the duration of quantum operations involving the ancilla $T_{\mathrm{gate}} \propto \chi_{gf}^{-1}$ is inversely proportional to the strength of the dispersive interaction between the cavity and the ancilla, and therefore typically on the order of $T_{\mathrm{gate}} \approx 1\,\si{\micro\second}$. Erasure errors are caused by single-photon loss events in the cavities happening at an average rate $\bar{\kappa}$ in the cavity system, as well as decay and dephasing events in the ancilla and the finite measurement infidelity $\eta_{\mathrm{ij}}$ for distinguishing the ancilla states $i$ and $j$. In general, the measurement infidelities to distinguish different transmon pointer states are not identical \cite{SalElder2020}, which is why we consider two separate infidelities $\eta_{ge} = 10^{-2}$ and $\eta_{gf} = 10^{-4}$. A difference in the single-photon decay rates $\Delta \kappa = \kappa_{\mathrm{b}} - \kappa_{\mathrm{a}}$ causes a no-jump backaction slowly polarizing the dual-rail towards the cavity with the longer lifetime, but, in contrast to the pure dephasing at rate $\gamma_{\varphi} = \gamma_{\varphi,{\mathrm{a}}} + \gamma_{\varphi,{\mathrm{b}}}$, contributes only to second-order to the Pauli error rate. By construction of the gates, ancilla decay events during the gate cause a backaction on the cavity system, giving rise to bit flip and phase flip errors at approximately similar rates in the case that they are undetected. 
For the calculation of the error probabilities we assume $\kappa_{\mathrm{a}} = (0.5\,\si{\milli\second})^{-1}$, $\kappa_{\mathrm{b}} = (1\,\si{\milli\second})^{-1}$, $\gamma_{\varphi,{\mathrm{a}}} = (20\,\si{\milli\second})^{-1}$, $\gamma_{\varphi,{\mathrm{b}}} = (20\,\si{\milli\second})^{-1}$, $\Gamma_{1}^{ge} = (100\,\si{\micro\second})^{-1}$,  $\Gamma_{1}^{ef} = (100\,\si{\micro\second})^{-1}$, $\Gamma_{\varphi}^{gf} = (100\,\si{\micro\second})^{-1}$. Notably, because heating events are rare in the cavities and the ancilla, we are not considering them here, but heating is also detected in most cases.}
\setlength{\tabcolsep}{20pt}
 \renewcommand{\arraystretch}{1.5}
\begin{tabular}{l||c|c|c|c}
\toprule
Error process & \multicolumn{2}{c|}{Erasure errors} & \multicolumn{2}{c}{Pauli errors}  \\ \hline \hline
Single-photon loss  & $\Bar{\kappa} T_{\mathrm{gate}}$ & $10^{-3}$ & - & - \\
no-jump backaction  & - & - & $\left( \frac{1}{4}\Delta \kappa T_{\mathrm{gate}}\right)^2$ & $10^{-6}$ \\
cavity dephasing & - & - & $\gamma_{\varphi} T_{\mathrm{gate}}$ & $10^{-4}$ \\
ancilla decay  & $\Gamma_{1}^{ef} T_{\mathrm{gate}}$ &  $10^{-2}$ & $\Gamma_{1}^{ef}\Gamma_{1}^{ge}T_{\mathrm{gate}}^2$ & $10^{-4}$\\
undetected ancilla decay  & - & - & $\eta_{\mathrm{ge}} \Gamma_{1,ef} T_{\mathrm{gate}}$ & $10^{-4}$\\
ancilla dephasing  & $\Gamma_{\varphi}^{gf} T_{\mathrm{gate}}$ & $10^{-2}$ & $\left(\Gamma_{\varphi}^{gf} T_{\mathrm{gate}}\right)^2$ & $10^{-4}$\\
photon loss + ancilla dephasing  & - & - & $\bar{\kappa} \Gamma_{\varphi}^{gf} T_{\mathrm{gate}}^2$ & $10^{-5}$\\
undetected ancilla dephasing  & - & - & $\eta_{\mathrm{gf}} \Gamma_{\varphi}^{gf} T_{\mathrm{gate}}$ & $10^{-6}$\\
measurement infidelity  & $\eta_{\mathrm{gf}}$ & $10^{-4}$ & - & - \\
photon loss + meas. infid. & - & - & $\eta_{\mathrm{gf}} \bar{\kappa} T_{\mathrm{gate}}$ & $10^{-7}$ \\
\end{tabular}
\label{Tab_errors_gate}
\end{table*}

\subsection{Erasure conversion}
\label{SEC:Erasure_conversion}
A keystone of the proposed dual-rail scheme is that we can perform QND detection of leakage out of the logical subspace caused by loss or gain of a photon in the cavities. For QEC protocols, the knowledge about the exact location of the error enables us to convert these otherwise pernicious leakage events into erasure errors, which are much easier to correct than Pauli errors. Therefore, we refer to this leakage detection scheme as the erasure check.\\
\\
Our erasure check is based on measuring the joint photon-number parity of the two cavities in a single dual-rail qubit. The dual-rail subspace has odd joint-parity but gain or loss of a photon changes this to even. Hence, QND joint-parity measurements allow us to perform erasure checks without destroying the logical states. Our specific scheme for measuring joint-parity is an extension of the well-established parity measurement scheme for an individual cavity\,\cite{Vlastakis_2013,Sun_2014}, and relies on the dispersive interaction to a single transmon ancilla as shown in Fig.\,\ref{fig:erasure_check}. \\
\\
Remarkably, by combining a beamsplitter interaction between the two dual-rail cavities with manipulations of a transmon coupled to only one cavity, there are simple procedures\,\cite{TakaGates2022} for measuring joint properties of the dual-rail qubit, without requiring precise matching of fabrication parameters. For the joint-parity erasure check, we perform a Ramsey sequence on the transmon with a wait time $T = 2 \pi/\chi_{gf}$ between the transmon $\pi/2$ pulses. During this time, the dispersive interaction between the transmon and one of the cavities maps photon-number information to the ancilla state, which is then measured at the end of the sequence. By also activating a beamsplitter interaction during the wait time, excitations swap back and forth between the cavities, allowing the ancilla to measure properties of both cavity states. \\
\\
We find that for a particular beamsplitter strength $g_{\mathrm{bs}} = \sqrt{3} \chi_{gf} / 2$ and detuning $\Delta = \chi_{gf}/2$ from resonance, we measure the joint-parity information of both cavities, which return to their initial states at the end of the sequence (up to a deterministic cavity phase shift that is easily tracked in software). In experiment, matching these conditions is easily achieved by adjusting the amplitude and detuning of microwave drives applied to the parametric coupler. These conditions are derived in App.\,\ref{APP:Section:CJP} and explored further in Tsunoda, Teoh et al.\,\cite{TakaGates2022}. In contrast to the joint-parity measurement implemented in Ref.\,\cite{Chen2016}, our method allows us to reserve the $\ket{e}$ level to detect transmon decay events, and only requires the ancilla to be coupled to one cavity.\\
\\
The purpose of the erasure check is to convert cavity leakage errors to erasure errors. However, the erasure check itself is error-prone. Of particular concern are transmon decay errors that happen with probability $\sim10^{-2}$ during the check. Left undetected, these errors induce cavity dephasing --- equivalent to a logical Pauli error on our dual-rail qubit. To combat this, we operate the transmon in the g-f-manifold and perform three-state readout\,\cite{SaclayfStateReadout2009,MartinisReadout2016,SalElder2020,WallraffReset2018,HuardReset2013} that distinguishes among the states $\ket{g},\ket{e}$ and $\ket{f}$. Measuring $\ket{e}$ now flags transmon decay events which are also converted to qubit erasures. Measuring $\ket{f}$ signals that we have leaked out of the dual-rail subspace.\\
\\
With this modification, the cost associated with each erasure check is an additional erasure probability on the $\sim 10^{-2}$ level and additional Pauli errors on the $\sim 10^{-4}$ level, due to second-order errors in the transmon. Since the construction of the erasure check is transparent to transmon dephasing\,\cite{MaPathIndependent2020}, we also see that transmon dephasing errors increase the erasure probability at the $\sim 10^{-2}$ level (see Table\,\ref{Tab_errors_gate}) by randomizing the even/odd measurement outcomes, hence representing a false positive leakage detection event. When a qubit passes the erasure check despite being in a leakage state we suffer from a false negative leakage detection event similar to a dark count in quantum optics. These errors will go on to cause Pauli errors in subsequent gate operations until they are most likely detected in the next round of leakage detection. These errors are still effectively second-order since they require both photon loss and a failed erasure conversion in-between rounds.

\begin{figure}[t]
	\includegraphics[width=.9 \linewidth] {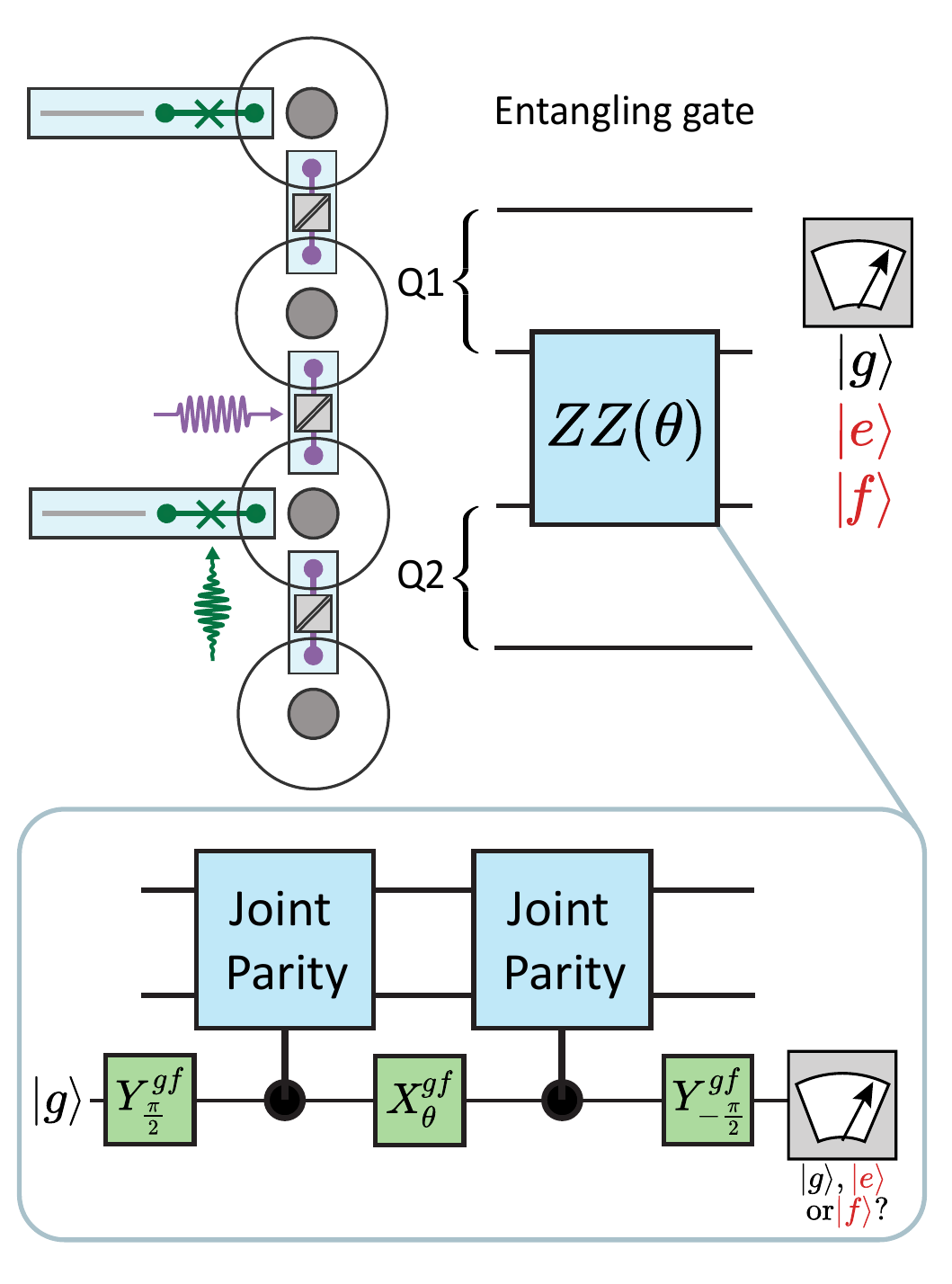}
	\caption{ 
        Hardware implementation of two cavity dual-rail qubits based on 3D superconducting cavities. The two-qubit entangling gate can be realized by a $ZZ(\theta)$ gate performed between the central rails only, for which only one additional coupler is required. For the proposed architecture, the $ZZ(\theta)$ gate can be implemented with the same building blocks as the erasure check, and is part of a larger family of error-detectable bosonic gates\,\cite{TakaGates2022}. Crucially for our architecture, transmon errors during the gate are detectable and mapped to flag states ($\ket{e}$: transmon decay, $\ket{f}$: transmon dephasing), allowing for the conversion of these physical errors into erasure errors. With local rotations on the dual-rail qubits, the $ZZ(\theta = \frac{\pi}{2})$ can be transformed into a CNOT or a CZ gate\,\cite{Makhlin_2002,Schuch_2003}.  
}
    \label{fig:two_qubit_gate}
\end{figure}

\subsection{Two-qubit gates}
Next we describe how entangling gates between dual-rail qubits can be constructed so that both cavity and transmon errors can also be detected and converted to erasures. 

The key insight is that one needs access to only a single cavity of a dual-rail qubit to realize the logical Z operator of the dual-rail. By driving a beamsplitter interaction between a pair of cavities, one from each of the two dual-rail qubits (see Fig.\,\ref{fig:two_qubit_gate}), a single transmon can effectively act as a control on both dual-rail qubits. Specifically, our proposed two-qubit entangling gate is the $ZZ(\theta)$ gate, which can  be written as
\begin{equation}
     \begin{pmatrix}
1 &  &  &  \\
 & e^{i\theta} &  &  \\
 &  & e^{i\theta} &  \\
 &  &  & 1 
\end{pmatrix}
\label{eq:ZZthetaU1}
\end{equation}
acting on the two-qubit logical subspace.
When $\theta=\pi/2$ this is locally equivalent to a CNOT or CZ gate\,\cite{Makhlin_2002,Schuch_2003}. 
A $ZZ(\theta)$ gate acting on the subspace $\{\ket{00},\ket{01},\ket{10},\ket{11}\}$ as if each cavity were encoded in the Fock $\{\ket{0},\ket{1}\}$ code will also realize the $ZZ(\theta)$ for the dual-rail code, provided each of these cavities belongs to a different dual-rail qubit.
\\
\\
Our motivation for the $ZZ(\theta)$ gate implementation is that it can be performed with the exact same hardware required for the erasure checks discussed in the previous section. Though other gate implementations are possible with an ancilla dispersively coupled to both cavities, with this approach, we only need to modify the control pulse sequence (see App.\,\ref{APP:ZZ_construction} for derivation).\\
\\ 
Similar to the error detection of the erasure check, it is of vital importance to detect transmon errors that happen during the gate in order to preserve the error hierarchy. The construction presented in Fig.\,\ref{fig:two_qubit_gate} ensures first-order transmon errors are detected when we measure the transmon at the end of the gate, erasing both dual-rail qubits if the transmon is not in $\ket{g}$. Once again, this operation is designed such that Pauli errors are only introduced from second-order hardware errors, when two decoherence events happen during a single gate. The expected error scalings for the $ZZ(\theta)$ gate associated with (detectable) first-order errors and second-order errors are shown in Table\,\ref{Tab_errors_gate}.\\
\\
Photon loss during the gate is detectable via separate erasure checks on the two dual-rail qubits after each gate, but modifications can also be made to the $ZZ(\theta)$ gate pulse sequence to simultaneously perform an erasure check (see App.\,\ref{APP:ZZ_with_detection}) in which the transmon can be mapped to the $\ket{f}$ level if one of the input dual-rail qubits was in a leakage state outside the logical subspace. Master equation simulations with various decoherence mechanisms and higher-order nonidealities in the system Hamiltonian are explored in detail in\,\cite{TakaGates2022} and suggest $\sim 10^{-2}$ erasure probability and $\sim10^{-4}$ Pauli errors per two-qubit gate, which makes the proposed gate the ideal building block for scalable QEC codes that correct for both Pauli and erasure errors\,\cite{wu_2022}.

\begin{figure}[t!]
	\includegraphics[width=1.0 \linewidth] {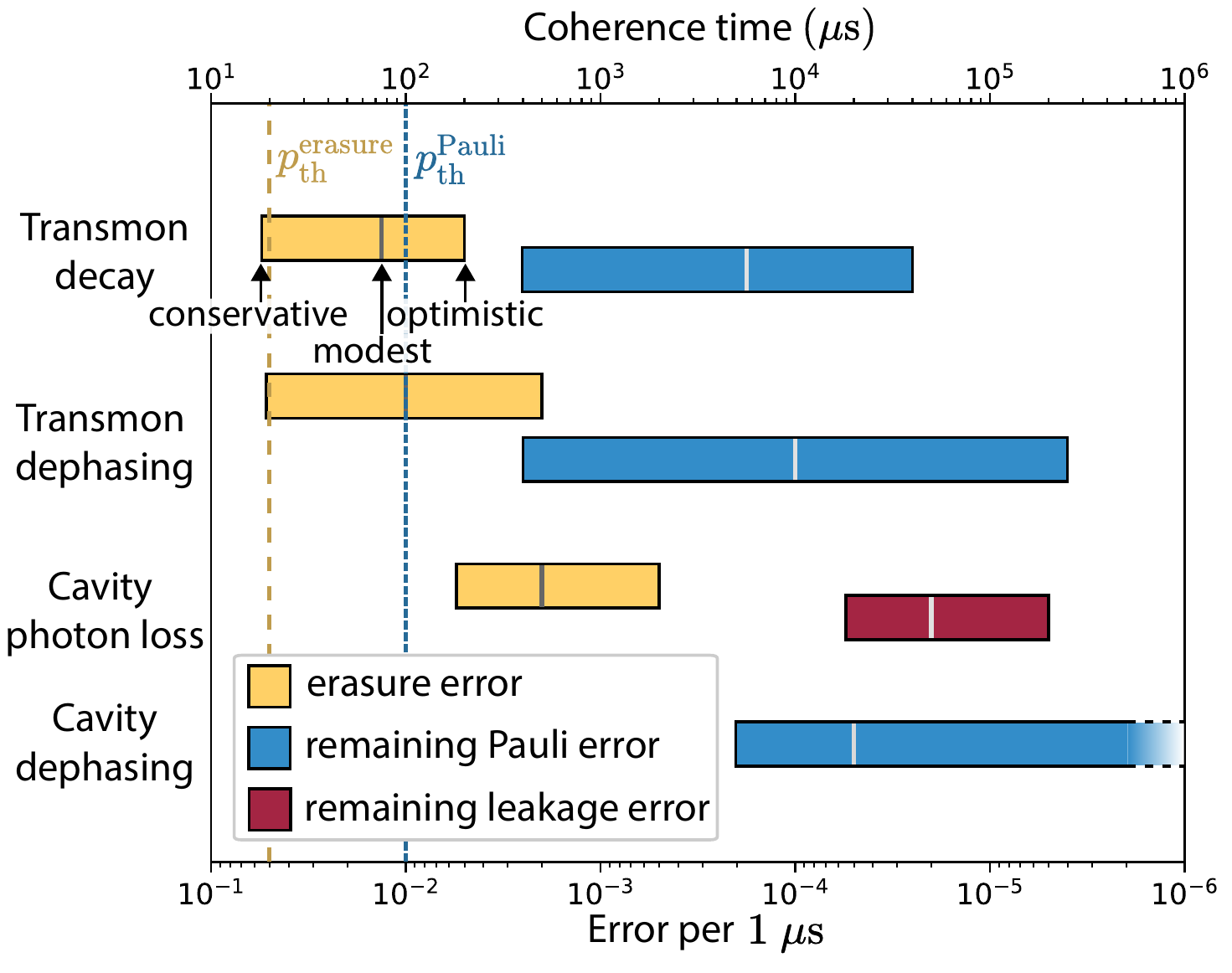}
	\caption{ 
Error hierarchy in the dual-rail encoding with superconducting cavities during operations. This is a visualization of the most important errors presented in Table\,\ref{Tab_errors_gate}. The most likely errors in the system are energy relaxation and pure dephasing in the ancilla transmon, followed by single-photon loss in the cavities. All of these errors can be converted into erasure errors to first order (yellow), leaving a significantly smaller, second-order floor of residual Pauli errors (blue) and leakage errors (dark red). The remaining first-order Pauli error is pure dephasing in the cavities, which is intrinsically low in superconducting cavities, especially when compared to all other error rates. For all physical error types we give a range for the error probability per $1\,\si{\micro\second}$ based on error rates typically found in the literature, from conservative (high error rate) to optimistic (low error rate). The values discussed in the main text are based on modest expectations for all components indicated by the colored lines. Remaining Pauli errors due to the transmons are estimated by squaring the transmon error rates and we assume a 99\% conversion efficiency for detecting single-photon loss, expected mainly due to transmon dephasing during joint-parity measurements. We also highlight the approximate thresholds for erasure and Pauli errors in a surface code architecture, $p_{\mathrm{th}}^{\mathrm{erasure}} = 5\%$ and $p_{\mathrm{th}}^{\mathrm{Pauli}} = 1 \%$, respectively, with vertical dashed lines, demonstrating that today's modest hardware is already expected to be well below threshold using our proposed dual-rail encoding.               
}
		\label{fig:DR_error_hierarchy}
\end{figure}

\begin{figure}[t!]
	\includegraphics[width=0.8 \linewidth] {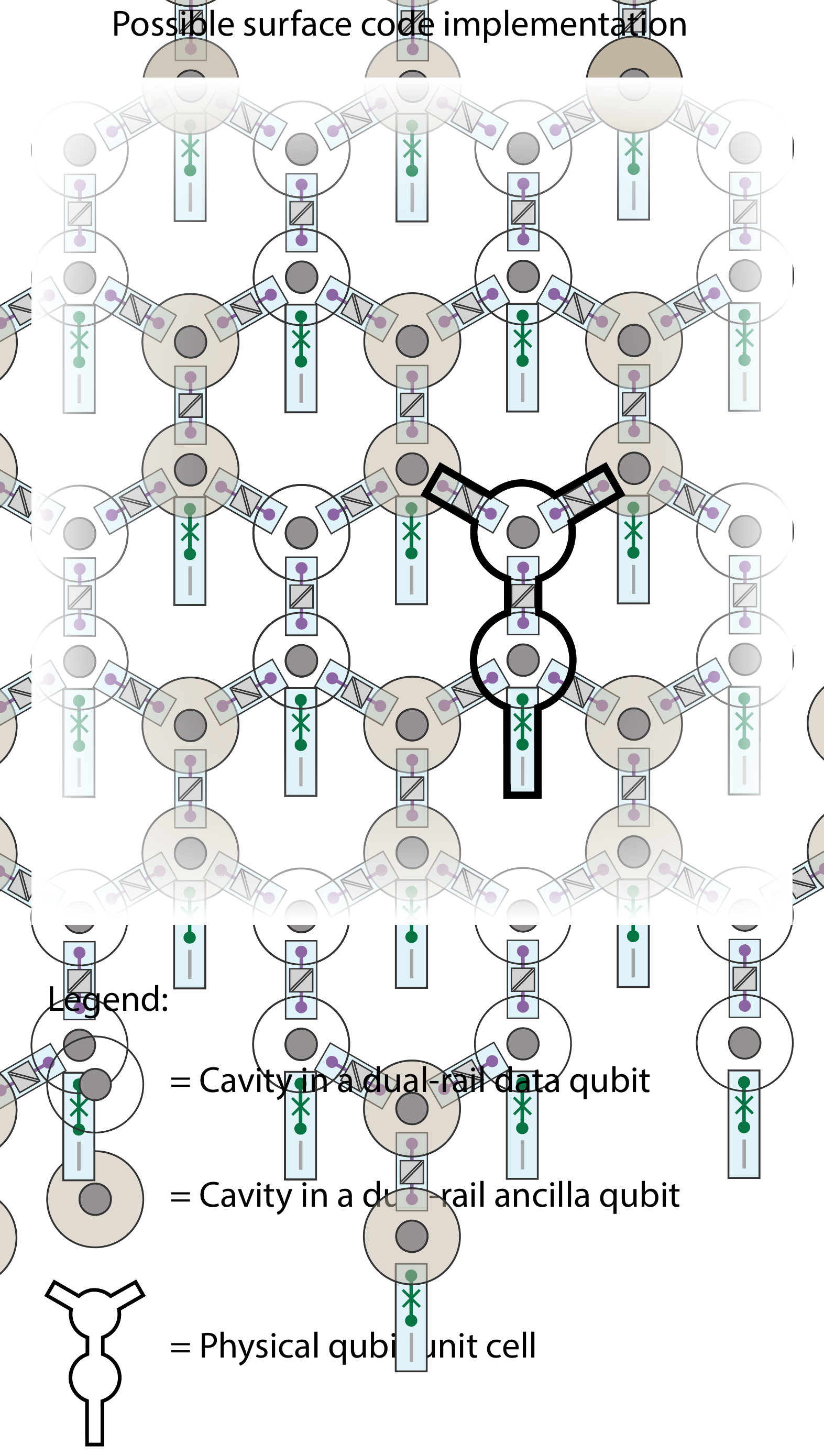}
	\caption{
Dual-rail cavity qubits forming the physical qubits in a lattice-based quantum error correction code. A single unit cell in the bulk of the lattice is outlined in black, and composed of two superconducting cavities (circles), a single ancilla transmon (green) with readout (grey), and three beamsplitter couplers (purple). Within a surface code, the nominally identical dual-rail qubits may play the role of either data qubits (white cavities), which store the quantum information, or measurement qubits (grey cavities), which are used to perform  stabilizer measurements during rounds of error correction. Nearest-neighbor connectivity between cavities is sufficient to build a 2D square lattice of dual-rail physical qubits, since entangling gates only require one cavity from each dual-rail qubit to interact together (see Fig.\,\ref{fig:two_qubit_gate}).  
}
		\label{fig:DR_surface_code}
\end{figure}

\section{Discussion}
\subsection{Reviewing the error hierarchy}
The principal feature of the proposed dual-rail qubit with superconducting cavities is its strong error hierarchy that persists throughout each operation. The error hierarchy is visualized for feasible coherence times in Fig.\,\ref{fig:DR_error_hierarchy}, ranging from conservative to optimistic hardware performance. Erasure errors result from first-order hardware errors and are predicted to occur with probability $\sim1\%$ per two-qubit gate. Pauli errors arise to first-order only from cavity dephasing, which is naturally suppressed in superconducting cavities, and from second-order hardware errors that are similarly rare, occurring with probability $0.01\%$ per two-qubit gate.  \\
\\
 Mitigating and containing leakage errors is an active field of research in superconducting quantum processors in general\,\cite{McEwen2021,Miao2022overcoming}. Even with finite erasure conversion efficiency, missed photon loss errors into the ground state are still detected with high probability in later rounds of erasure conversion, meaning a constant, small fraction ($\sim10^{-5}$ in equilibrium if we assume $10^{-3}$ probability of leakage and 99\% erasure conversion efficiency) of dual-rail qubits remain in $\ket{00}$. Attempts to perform entangling gates with a qubit in $\ket{00}$ result in an identity operation instead. At worst, these leaked states act as a source of Pauli errors on surrounding unleaked qubits whenever we attempt an entangling gate, until they are eventually detected and reset by re-initialization into the dual-rail subspace. Importantly, the fraction of leaked qubits is not expected to grow in time. All other forms of leakage are expected to be extremely rare but in principle can also be detected with the appropriate transmon measurements. \\
\\
These properties make dual-rail qubits immediately desirable for general quantum computing applications. For NISQ-era algorithms\,\cite{preskill_2018}, we can run short-depth circuits with post-selection, abandoning a circuit run if any erasures are detected, resulting in very low SPAM errors and high effective gate fidelity. The yield of successful circuit attempts decreases exponentially with circuit depth, but for the runs we keep, only very low Pauli error probabilities will remain.  With $1\%$ erasure probability we expect to run circuits with up to $\sim100$ two-qubit gates before the success probability becomes impractically low. We emphasize that dual-rail qubit Pauli errors scale quadratically with ancilla transmon lifetimes when operating the transmon in the g-f-manifold, whereas Pauli errors in a transmon-only architecture scale linearly. Thus we expect dual-rail qubit gate fidelities to improve faster than transmon qubit gate fidelities as transmon coherences continue to improve, whilst the transmon remains as the dominant error source.
\subsection{Scalability}
Perhaps the most exciting use of dual-rail cavity qubits is as the ‘physical qubits’ used to encode the logical qubits of a topological fault-tolerant error correcting code such as the surface code. Many QEC codes can tolerate erasure errors much better than Pauli errors\,\cite{GrasslErasure1997}, which is reflected in their respective error thresholds\,\cite{StaceSurface25Erasure2009,Delfosse2020,Wang_2011}.
While the exact threshold is highly dependent on the order of gate operations and frequency of error detection, for simplicity, we envision performing error detection (via joint-parity measurements) after every two-qubit $ZZ(\theta)$ gate. A general version of this noise model was previously studied for a neutral atoms platform\,\cite{wu_2022}, where the threshold for erasure errors was about 5 times higher than for Pauli errors, even with finite detection efficiency.  However, error detection after every two-qubit gate may not be the best strategy for the dual-rail encoding, but will likely depend on the hardware parameters and relative cost of performing error detection. 
\\
\\
We are also aware of the seemingly higher erasure thresholds that exist in measurement-based error correction schemes\,\cite{StaceSurface25Erasure2009}. However, the obtained thresholds are highly depend on the underlying error model and are actually similar to gate-based approaches under more realistic scenarios\,\cite{Whiteside_2014}. In general, a fair comparison with gate-based approaches based on only the thresholds requires detailed accounting of the number of steps needed to prepare and measure the resource states, as well as detailed knowledge of circuit-level noise, and is therefore beyond the scope of this manuscript. Undoubtedly, a measurement-based approach requires more physical qubits, an overhead which could otherwise be used to increase the code distance or number of logical qubits in a gate-based approach.\\
\\
Since a surface code of distance $d$ can tolerate up to $d-1$ erasure errors per error correction cycle but only $(d-1)/2$ Pauli errors, erasure errors not only have higher thresholds, but scale better with increasing code distance. To put this into perspective, for a fixed distance $d=11$, a surface code with only erasure errors at $\sim1\%$ per two-qubit gate will have roughly the same performance as the same surface code with only Pauli errors at $\sim0.05\%$ per gate, resulting in logical error rates $\sim10^{-8}$ per cycle. In general, when both Pauli and erasures are present, we need to ensure both error rates are substantially below their thresholds\,\cite{wu_2022}.
\\
\\
Despite tremendous efforts in realizing transmon-based surface codes, reaching physical error rates far below the Pauli threshold is still a major outstanding challenge\,\cite{Marques2022,Krinner2022,Google_EC_gain_2022}. We believe our approach offers a compelling alternative, whereby error detection opens a shortcut to achieving error rates far below the relevant thresholds, without the need to drastically increase coherence times.\\
\\
The favorable scaling of erasure errors with distance indicates that detecting first-order hardware errors and flagging them as erasures is almost as good as fully correcting them at the hardware level. Both approaches scale the same with code distance, $d$ and differ only by a prefactor that characterizes the respective overheads. 
Rather than correcting quantum errors at the hardware level, we now error detect at the hardware level and reset the erased qubits, a simpler task overall.
\\
\\
\\
\\
Finally, scaling up with a 3D cavity dual-rail architecture has the potential to greatly suppress unwanted cross-talk and package modes in a multiqubit processor, as well as correlated errors between the physical qubits. A conceptual illustration of a 2D square lattice of dual-rail qubits required to implement a topological quantum error correction code is shown in Fig.\,\ref{fig:DR_surface_code}. The fact that we can interact with a dual-rail qubit through either one of the cavities means we can reduce the co-ordination number for each cavity mode in lattice-based QEC codes, while maintaining nearest-neighbor connectivity among the dual-rail qubits. The unit cell of our lattice consists of a dual-rail qubit, comprising two superconducting cavity modes, a transmon with readout and a switchable beamsplitter coupler between the cavity modes, and two additional beamsplitter couplers to interface with neighbouring unit cells. In comparison, a transmon-based surface code may have unit cells consisting of a single transmon and two coupling elements. \\
\\
In general, increasing the code distance of a surface code exponentially suppresses the logical error rate, provided physical error rates remain well below threshold when increasing the number of hardware components. Alternatively, one can focus on reducing the physical error rate further below threshold, which gives a similar exponential improvement. In the dual-rail encoding, we are willing to increase the complexity of our physical qubit in return for significantly lower physical error rates, which now also reduce quadratically with improvements in transmon coherence times \,\cite{TakaGates2022}. When we then scale up to larger code distances, this likely results in far fewer hardware components to reach a target logical error rate, and allows an additional margin for sources of errors associated with scaling up any qubit architecture.

\section{Conclusion}
We have introduced the dual-rail cavity qubit for circuit-QED, which leverages the intrinsic noise bias of microwave cavities and the dual-rail code to make a fully error-detected logical qubit. In this paradigm, we detect and convert the dominant first-order hardware errors to erasure errors, leaving a small undetected background of second-order errors which become Pauli errors. We expect erasures per gate at the $\sim 1\%$ level and Pauli errors at the $\sim 0.01\%$ level for typical present-day coherence times, making both values well below their respective thresholds of $\sim 5\%$ and $\sim 1\%$, respectively. Furthermore, all leakage out of the logical subspace can in principle be detected. Most importantly, this strong hierarchy of errors can be maintained while performing single and two-qubit gates, thereby relaxing the requirements for quantum error correction, and the realization of a fault-tolerant quantum computer. Each dual-rail qubit can play the role of a physical qubit in a surface code, where erasures are significantly easier to correct than Pauli errors, having both a higher threshold and superior scaling with code distance. \\
\\
Our results suggest that the dual-rail code is the most efficient bosonic encoding for microwave cavities, benefiting greatly from straightforward single-qubit gates, measurement and state preparation whilst incurring minimal photon number overhead relative to other bosonic codes. Realizing all logical operations is imminently achievable, with all hardware requirements already demonstrated. With today’s typical coherence times, we anticipate being significantly below the effective threshold for a surface code with erasure and Pauli errors present. Since this approach makes all physical errors second-order at the hardware level, we also predict that the performance of dual-rail qubits will improve more rapidly than conventional schemes as coherence times improve further. Finally, this paradigm can find immediate use in implementing near-term short-depth circuits with post-selection to improve the utility of today's NISQ applications. 

\section{Materials and Methods}
For the simulation results in Fig.\,\ref{fig:state_prep_measurement}, we model transmon readout by considering two dominant sources of readout error. This is decay of the transmon during the readout pulse and the finite separation of the transmon's pointer states with the full error model detailed in App.\,\ref{sec:msmt_model}. Further details on how Lindblad master equation simulations were used to model the entire logical readout protocol can be found in App.\,\ref{sec:msmt_strat}. We can also adjust the measurement decoding scheme to either favour low erasure rates or low logical assignment errors.
\pagebreak

\section{Acknowledgement}
We thank Yue Wu for helpful discussions and simulations for a surface code QEC with both erasure and Pauli errors. \\
\\
This research was supported by the U.S. Army Research Office (ARO) under grants W911NF-18-1-0212, W911NF-16-1-0349 and W911NF-22-1-0053, and by the U.S. Department of Energy, Office of Science, National Quantum Information Science Research Centers, Co-design Center for Quantum Advantage (C2QA) under contract number DE-SC0012704. The views and conclusions contained in this document are those of the authors and should not be interpreted as representing official policies, either expressed or implied, of the ARO or the U.S. Government. The U.S. Government. is authorized to reproduce and distribute reprints for Government purpose notwithstanding any copyright notation herein. 
\\
\\
RJS and LF are founders and shareholders of Quantum Circuits, Inc. SMG, SP are equity holders of Quantum Circuits, Inc.

\bibliography{main}

\begin{titlepage}\centering
\vspace*{\fill}
\LARGE \textbf{Supplementary Information}
\vspace*{\fill}
\end{titlepage}

\appendix
\section{Modelling transmon readout errors}
\label{sec:msmt_model}
Many operations in our dual-rail scheme (logical measurement, erasure checks, error-detected entangling gates)  rely on high fidelity single shot readout of a transmon ancilla. In our analysis, we take into account errors that are experimentally known to occur during this readout, and show they do not significantly affect logical performance. We consider decay of the transmon during the readout and readout misassignment due to the finite signal-to-noise ratio (SNR) of the readout signal. For numerical simulations of logical measurement, we model transmon readout in the $g$-$e$ manifold as follows:
First we calculate the (unnormalized) density matrices for the cavities, $\rho_g^c$ and $\rho_e^c$ that would result from perfect transmon measurement. We obtain these from
\begin{align*}
\rho_g &= \ket{g}\bra{g}\rho\ket{g}\bra{g}\\
\rho_e &= \ket{e}\bra{e}\rho\ket{e}\bra{e},
\end{align*}
where $\rho$ is the combined state of the transmon and cavities before the measurement. We obtain $\rho_g^c$ by tracing over the ancilla state in $\rho_g$ and similarly for $\rho_e^c$. Given a particular transmon measurement outcome, $M=\{g,e\}$, we model the state of the system to be
\begin{align*}
    &\rho_{M=g} =\\
    &\left((1-P_{\mathrm{o}})\rho_g^c + \frac{P_{\mathrm{d}}}{2}\rho_e^c\right)\otimes\ket{g}\bra{g} + P_{\mathrm{o}}\rho_e^c\otimes\ket{e}\bra{e}\\
&\rho_{M=e} =\\
&(1-P_{\mathrm{o}}-P_{\mathrm{d}})\rho_e^c\otimes\ket{e}\bra{e} +\left( \frac{P_{\mathrm{d}}}{2}\rho_e^c+P_{\mathrm{o}}\rho_g^c\right)\otimes\ket{g}\bra{g}.  
\end{align*}\\
We set $P_{\mathrm{d}}=0.01$ to model a 1\% chance of transmon decay at a random time during the readout. When this happens, half the time we will record the transmon in $\ket{g}$ and half the time we record $\ket{e}$. The overlap error $P_{\mathrm{o}}=10^{-4}$ is the probability of misassigning $\ket{g}$ as $\ket{e}$ and vice versa due to noise in our readout signal and is representitive of current state-of-the-art transmon readout \cite{WallraffReadout2017}. Additionally, cavity decay during the readout is modelled by including a $1 \si{\micro\second}$ idle time for a typical readout duration. This model can easily be extended to include the $\ket{f}$ level but is not necessary for logical readout simulations. Another advantage when using a three-level transmon for the erasure check and entangling gate is that transmon decay from $\ket{f}$ to $\ket{e}$ during transmon readout will be flagged as erasure whereas two decay errors are necessary to cause a Pauli error.

\section{Numerical simulations for Dual-rail logical measurement}
\label{sec:msmt_strat}
 
\begin{figure}[h]
	\includegraphics[width=\linewidth] {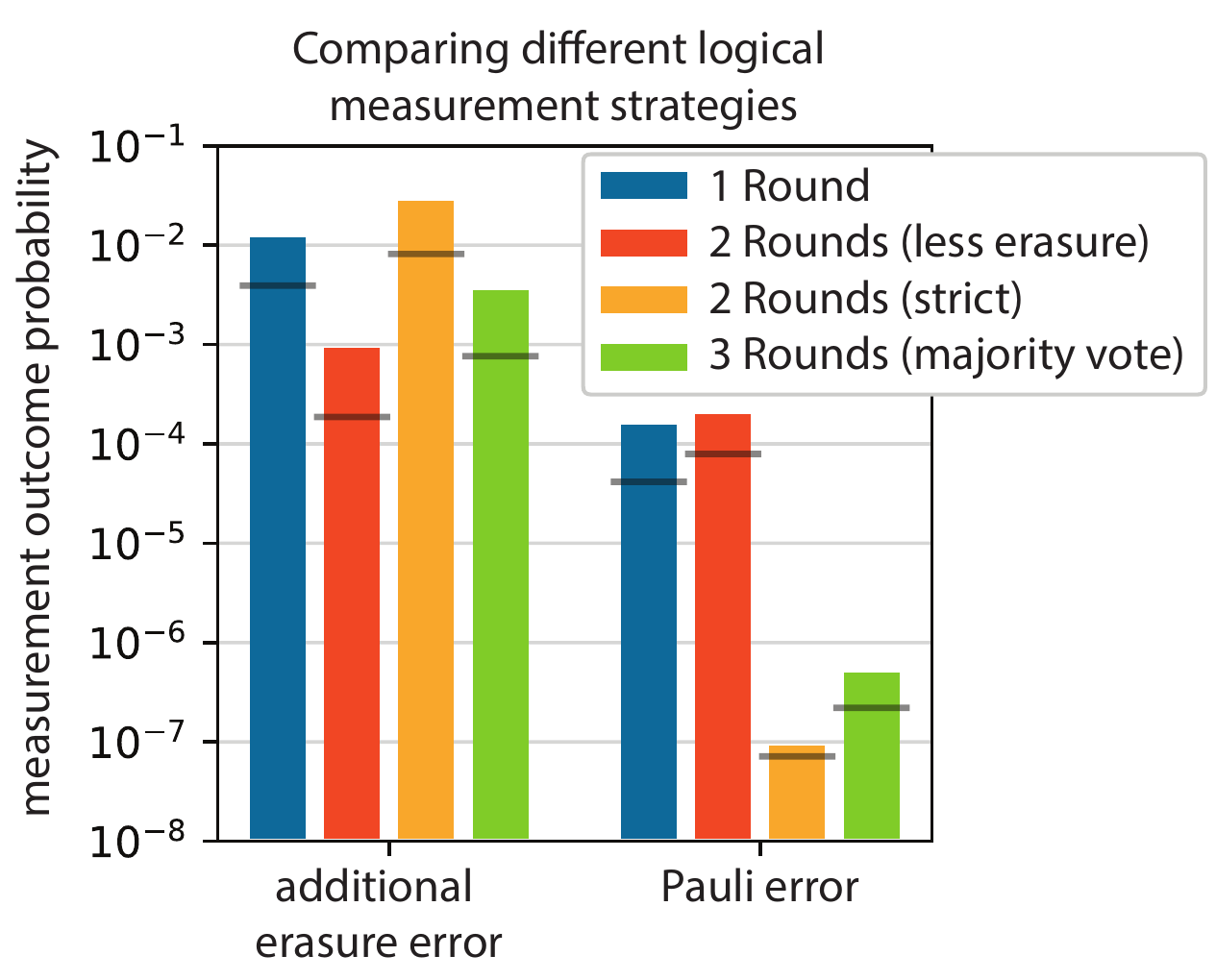}
	\caption{Comparison of 4 possible dual-rail logical measurement schemes listed in Table IV for $p=0.01$ and coherence times listed in Table III. The horizontal grey bars indicate the logical measurement performance if transmon readout were perfect. By changing our strategy, we can minimize additional erasure errors from the measurement or Pauli error. By majority voting, we can in principle reach exceedingly low Pauli rates. In experiment, this may be limited by our ability to keep the transmon in the g-e-manifold throughout repeated measurements.}
		\label{fig:Sfig_msmt}
\end{figure}

Here we descirbe how the QuTiP Lindblad Master equation simulations were performed to estimate logical measurement fidelity for a dual-rail cavity qubit under realistic hardware coherence times listed in Table\,\ref{Tab_coh_times}. For each round of measurement we first prepare the cavities in either $\ket{01},\ket{10}$ or $\ket{00}$ and the transmon in $\ket{g}$. We then perform the standard parity mapping sequence, $\pi/2$ pulse --- wait time $\pi/\chi$ --- $\pi/2$ pulse, in order to map parity information to the state of the transmon \cite{Sun_2014}. We then simulate imperfect transmon measurement according to the readout model. For $M=e$ outcomes, we perform a perfect transmon $\pi$-pulse to reset the transmon. We then apply a perfect SWAP between the cavity modes and repeat the whole parity measurement sequence once more to obtain another transmon measurement outcome. The measurement round is concluded by another perfect SWAP operation to return the states to their initial cavities. \\
\\
In one round of measurement we may obtain one of four outcomes (g,g),\,(g,e),\,(e,g) or (e,e). We store the probability and density matrix for each of these outcomes. 
Every time we increase the number of measurement rounds, the total number of outcomes and stored density matrices grows by a factor of 4. We group all possible outcome strings into three bins: declare $\ket{0}_{\mathrm{L}}$, declare $\ket{1}_{\mathrm{L}}$ or declare erasure. By summing the probabilities of each of the outcomes in a given bin, we calculate the probability of declaring a logical outcome for a given input state. \\
\\
We take $p=0.01$, the probability that the  cavities are in $\ket{00}$ at the start of the measurement, and represents the probability of a prior leakage error (after $\sim10$ entangling gates). Incorrectly declaring a leakage state as a valid logical state is a source of Pauli error included in the simulations. By changing the number of rounds, and the binning of our outcome strings, we can change our `measurement strategy' to favor lower Pauli errors but higher additional erasure errors. `False positive' erasure errors mean we reset qubits that did not need to be reset, resulting in an overall increase in our erasure rate.\\
\\
In Fig.\,\ref{fig:state_prep_measurement}, we compare the 1-round measurement strategy with a 3-round strategy with majority voting. In Table\,\ref{Tab_msmt_outcomes} we describe the outcome binning for these two strategies as well as two other strategies for 2-round measurement. One of which favors reducing the additional erasure rate incurred by the logical measurement and another which favors low Pauli error but higher erasure rates (`strict'). Simulated logical performance for all four measurement strategies is shown in Fig.\,\ref{fig:Sfig_msmt} and demonstrates the measurement's robustness to transmon readout errors. 

\begin{table}[h]
\label{tab:sim_coherence}
\begin{tabular}{c|c}
Collapse Operator & Coherence Time \\ \hline
$\ket{g}\bra{e}$               & 100 $\mu$s         \\
$\ket{e}\bra{e}$                & 100 $\mu$s         \\
$a,b$                 & 1 ms             \\
$a^\dagger,b^\dagger$                & 100 ms           \\
$\ket{e}\bra{g}$                & 10 ms            \\
\end{tabular}
\caption{Collapse operators and corresponding decoherence times used in Linblad master equation simulations for logical measurement. In all simulations, $\chi_{ge}/2\pi=-2 \text{ MHz}$. Cavity dephasing is not included since this does not affect logical readout.
}
\label{Tab_coh_times}
\end{table}
\begin{table}[h]
\label{tab:outcome_bins}
\begin{tabular}{l||c|c|c}
\toprule
& \multicolumn{3}{c}{\textbf{Declared Outcome}}\\
& $\ket{0}_{\mathrm{L}}$ & $\ket{1}_{\mathrm{L}}$ & Erasure \\ \hline\hline
1R & (e,g) & (g,e) & (g,g),(e,e)  \\ \hline
\begin{tabular}[c]{@{}c@{}}2R\\ \end{tabular}   & \begin{tabular}[c]{@{}c@{}}(e,g,-,-)\\ (g,g,e,g),(e,e,e,g)\end{tabular} & \begin{tabular}[c]{@{}c@{}}(g,e,-,-)\\ (g,g,g,e),(e,e,g,e)\end{tabular}& \begin{tabular}[c]{@{}c@{}}(g,g,g,g),(e,e,e,e)\\ (g,g,e,e),(e,e,g,g)\end{tabular}\\ \hline
\begin{tabular}[c]{@{}c@{}}2R$^*$\\ \end{tabular}&(e,g,e,g)&(g,e,g,e)&all other strings  \\ \hline
\begin{tabular}[c]{@{}c@{}}3R \\ \end{tabular} & ($\tilde{e},\tilde{g}$)&($\tilde{g},\tilde{e}$)& ($\tilde{g},\tilde{g}$),($\tilde{e},\tilde{e}$)  \\
\end{tabular}
\caption{Outcome binning for different logical measurement strategies. 1R --- 1-round measurement, 2R---2-round measurement favoring less erasure, 2R$^*$---2-round measurement (strict), 3R---3-round measurement with majority voting. Outcomes are listed chronologically left to right with two outcomes per measurement round. `-' denotes the value can be either g or e, $\tilde{g} = \{(g,g,g),(g,g,e),(g,e,g),(e,g,g,)\}$, the majority vote in favor of $g$ (for measurements on the same cavity). $\tilde{e} = \{(e,e,e),(e,e,g),(e,g,e),(g,e,e,)\}$, majority vote in favor of $e$ .
}
\label{Tab_msmt_outcomes}
\end{table}

\newpage
\section{Logical subspace and single-qubit operations}
The logical code words and eigenstates of the logical Pauli $Z$ operator in the dual-rail subspace are the single-photon Fock states
\begin{align}
    \ket{0}_{\mathrm{L}} &:= \ket{01} \\
    \ket{1}_{\mathrm{L}} &:= \ket{10}.
\end{align}
Hence, an arbitrary logical state of the dual-rail qubit expressed in the Fock basis of the combined cavity system is 
\begin{equation}
    \ket{\Psi}_{\mathrm{L}} = u \ket{01} + v\ket{10},
\end{equation}
where $u$ and $v$ are generally complex probability amplitudes satisfying $|u|^2 + |v|^2 = 1$ for a pure state. In the following, we will demonstrate how logical operations in the dual-rail subspace can be implemented with a simple beamsplitter interaction between the cavities. The photon-number preserving beamsplitter Hamiltonian is 
\begin{align}
    \mathcal{H}_{\mathrm{bs}} / \hbar &= \frac{g(t)}{2} a b^\dagger + \frac{g^\star(t)}{2} a^\dagger b \\
        &= \frac{g_{\mathrm{bs}}(t)}{2} \left(a b^\dagger e^{- i \phi(t)} + a^\dagger b e^{i \phi(t)} \right),
\label{APP:EQ_beamsplitter_simple}
\end{align}
where $a^\dagger$, $b^\dagger$ and $a$, $b$ are the single-mode creation and annihilation operators of the two linear modes denoted \textit{Alice} and \textit{Bob} in the main text, respectively, and $g(t)$ is the time-dependent, generally complex beamsplitter rate. Experimentally, the indicated time-dependence of the beamsplitter rate $g_{\mathrm{bs}}(t)$ is controlled by the amplitude of the drive tone(s) applied to the parametric coupler, while the phase $\phi(t)$ is determined by the phase of the drive tone(s) and can be updated in software. We can diagonalize the beamsplitter Hamiltonian given in Eq.\,\ref{APP:EQ_beamsplitter_simple} by means of a unitary coordinate transformation of the form
\begin{equation}
\begin{pmatrix}
d_1 \\
d_2 \\ 
\end{pmatrix}
= \frac{1}{\sqrt{2}}
\begin{pmatrix}
1 & - e^{i \phi} \\
e^{-i\phi}  &  1 \\
\end{pmatrix}
\begin{pmatrix}
a \\
b \\ 
\end{pmatrix},
\label{APP:beamsplitter_dressed_basis_general}
\end{equation}
where $d_1$ and $d_2$ are the single-mode annihilation operators of the dressed basis. The transformed Hamiltonian is
\begin{equation}
\mathcal{\tilde{H}} / \hbar = - \frac{g_{\mathrm{bs}}(t)}{2} d_1^\dagger d_1 + \frac{g_{\mathrm{bs}}(t)}{2} d_2^\dagger d_2
\end{equation}
with the corresponding eigenvectors 
\begin{equation}
\ket{\Psi_\pm} = \frac{1}{\sqrt{2}}\left(\ket{01} \pm e^{i \phi} \ket{10} \right)
\end{equation}
 associated to the eigenenergies $E_{\pm} = \pm g_{\mathrm{bs}} / 2$. In order to calculate the time-evolution of the dual-rail code words under the action of the beamsplitter Hamiltonian, we can express both states in the dressed basis:
\begin{align}
\ket{01} &= \frac{1}{\sqrt{2}}\left(\ket{\Psi_+} + \ket{\Psi_-}\right) \\
\ket{10} &= \frac{1}{\sqrt{2}}\left(\ket{\Psi_+} - \ket{\Psi_-}\right) e^{-i \phi}
\end{align}
Since the time-evolution of the states $\ket{\Psi_\pm}$ is determined by their associated eigenenergies 
\begin{equation}
U(t) \ket{\Psi_\pm} = e^{\mp i g_{\mathrm{bs}} t / 2}\ket{\Psi_{\pm}},
\end{equation}
we can express the time evolution operator $U(t)$ in the dual-rail subspace as a function of the beamsplitter rate $g_{\mathrm{bs}}$ and the phase $\phi$:
\begin{equation}
U(t) = 
\begin{pmatrix}
\cos\left(\frac{g_{\mathrm{bs}} t}{2}\right) & - i e^{- i \phi} \sin\left(\frac{g_{\mathrm{bs}} t}{2}\right) \\
- i e^{i \phi} \sin\left(\frac{g_{\mathrm{bs}} t}{2}\right) & \cos\left(\frac{g_{\mathrm{bs}} t}{2}\right)
\end{pmatrix}
\label{APP:EQ:unitary_evolution_dual_rail}
\end{equation}
By introducing the polar angle $\theta = g_{\mathrm{bs}} t$ and interpreting the phase $\phi$ as the azimuthal angle, we can identify the action of the beamsplitter interaction as an arbitrary rotation in the dual-rail subspace.
\begin{equation}
U(\theta, \phi) = 
\begin{pmatrix}
\cos\left(\frac{\theta}{2}\right) & - i e^{- i \phi} \sin\left(\frac{\theta}{2}\right) \\
- i e^{i \phi} \sin\left(\frac{\theta}{2}\right) & \cos\left(\frac{\theta}{2}\right)
\end{pmatrix}
\label{APP:EQ:unitary_rotation}
\end{equation}
\subsection*{SWAP operation}
For $\theta = \pi$ and $\phi = \pi$, we realize a logical \textit{SWAP} operation up to the rotation operator $e^{i\frac{\pi}{2}\left(\hat{a}^\dagger\hat{a}+\hat{b}^\dagger\hat{b}\right)}$, which is a global phase for a dual-rail qubit. A SWAP corresponds to the logical $X$ operator.
\subsection*{Hadamard operation}
For $\theta = \frac{\pi}{2}$ and $\phi = \pi$ we realize a logical \textit{Hadamard} operation, $H_{12}$ up to a phase rotation of the first cavity
\begin{equation}
    U(\theta = \frac{\pi}{2}, \phi = \pi) = S_1 \otimes \mathbb{1}_2 \times H_{12} \times S_1 \otimes \mathbb{1}_2.
\end{equation}
Here, we have used that the phase gate on the first cavity is
\begin{equation}
S_1 \otimes \mathbb{1} = 
\begin{pmatrix}
1 & 0 \\ 
0 & i
\end{pmatrix}
\end{equation}
in the dual-rail subspace. 

\section{The no-jump backaction and pure dephasing in the dual-rail subspace}
\label{App:no_jump}
As we emphasize in the main text, the dominant physical error channel in superconducting cavities is single-photon loss at rate $\kappa_i$, where the index $i \in \{\mathrm{a},\mathrm{b}\}$ indicates the respective mode in a single dual-rail qubit. Assuming we can perfectly detect these events, the remaining errors are pure dephasing of both cavities due to frequency fluctuations occurring at rates $\gamma_{\varphi, \mathrm{a}}$ and $\gamma_{\varphi, \mathrm{b}}$, and a no-jump backaction in case the loss rates of the two cavities into the common ground state are not identical $\Delta \kappa = \kappa_{\mathrm{b}} - \kappa_{\mathrm{a}}\neq 0$. \\
\\
We can calculate the effect of both processes based on a Krauss representation of the associated quantum channels. According to Micheal et al. \cite{MichaelBin2016}, the Krauss operator associated with the no-jump backaction is
\begin{equation}
    E_{00} = E_{00}^\dagger = e^{-\frac{1}{2}\left(\kappa_{\mathrm{a}} a^\dagger a + \kappa_{\mathrm{b}} b^\dagger b \right) t},
\end{equation}
and consequently photon-number preserving. Since the process of pure dephasing is also photon-number preserving and assuming a temperature $T = 0$, we can restrict the discussion to the dual-rail subspace $\{\ket{01}, \ket{10}\}$, ignoring heating events from the common ground state back into the dual-rail subspace. We can calculate the time-evolution due to the no-jump backaction for an arbitrary initial state $\ket{\Psi_0}$ 
\begin{equation}
\rho(0) = \ket{\Psi_0}\bra{\Psi_0} = 
\begin{pmatrix}
 |v|^2 & u^\star v \\ 
  u v^\star & |u|^2 \\ 
\end{pmatrix}
\label{init}
\end{equation}
by calculating  
\begin{equation}
    \rho(t) =  \frac{E_{00} \rho (0) E_{00}^\dagger}{|\bra{\Psi_0} E_{00}^\dagger E_{00} \ket{\Psi_0}|}.
\end{equation}
Here, the denominator ensures the normalization of the density matrix at every time increment. In addition, the pure dephasing of both cavities act as phase damping channels in the dual-rail subspace
\begin{align}
    E_{0} &=  \begin{pmatrix}
        \sqrt{1 - p_{\mathrm{b}}} & 0 \\
        0 & \sqrt{1 - p_{\mathrm{a}}}
    \end{pmatrix} \\
    E_{1} &=  \begin{pmatrix}
        \sqrt{p_{\mathrm{b}}} & 0 \\
        0 & 0
    \end{pmatrix} \\
    E_{2} &= \begin{pmatrix}
        0 & 0 \\
        0 & \sqrt{p_{\mathrm{a}}}
    \end{pmatrix},
\end{align} 
where the probabilities are  $p_i = 1 - e^{-2\gamma_{\varphi,i} t}$, with $i \in \{\mathrm{a,b}\}$, leading to an exponential decay of the off-diagonal terms in the density matrix at rate $\gamma_{\varphi} = \gamma_{\varphi, \mathrm{a}} + \gamma_{\varphi, \mathrm{b}}$. The analytical solution is 
\begin{equation}
\rho(t) =
\begin{pmatrix}
\frac{|v|^2 e^{-\Delta \kappa t}}{1 - |v|^2 \left(1 - e^{-\Delta \kappa t} \right)}  & \frac{u^\star v e^{- \frac{1}{2} \Delta \kappa t}}{1 - |v|^2 \left(1 - e^{-\Delta \kappa t} \right)} e^{- \gamma_\varphi t}\\ 
\frac{u v^\star e^{- \frac{1}{2} \Delta \kappa t}}{1 - |v|^2 \left(1 - e^{-\Delta \kappa t} \right)}e^{- \gamma_\varphi t} & \frac{|u|^2}{1 - |v|^2 \left(1 - e^{-\Delta \kappa t} \right)}\\ 
\end{pmatrix}.
\label{rho_no_jump}
\end{equation}
We can see immediately, that the no-jump backaction vanishes in case the single-photon decay rates are identical ($\Delta \kappa = 0$), leaving the system in the initial state. In case there is a finite difference in the single-photon decay rates, the time-dependent expectation values of the logical Pauli operators are 
\begin{align}
\braket{X(t)} &= \frac{\left(u v^\star + u^\star v   \right) e^{- \frac{1}{2} \Delta \kappa t}}{1 - |v|^2 \left(1 - e^{-\Delta \kappa t} \right)}e^{- \gamma_\varphi t} \\
\braket{Y(t)} &= \frac{\mathrm{i} \left( u v^\star - u^\star v  \right) e^{- \frac{1}{2} \Delta \kappa t}}{1 - |v|^2 \left(1 - e^{-\Delta \kappa t} \right)}e^{- \gamma_\varphi t}\\
\braket{Z(t)} &= \frac{|u|^2 - |v|^2 e^{-\Delta \kappa t}}{1 - |v|^2 \left(1 - e^{-\Delta \kappa t} \right)}
\label{pred_Pauli_t}
\end{align}
Consequently, the two code words are stable since the time-dependence of $\braket{Z}$ vanishes for $|u|^2 = 1$ and $|v|^2 = 1$. Hence, only superposition states with finite overlap to both code words will be polarized towards the cavity with lower decay rate ($\Delta \kappa$ can be positive or negative). \\
\\
We can calculate the average state fidelity as a function of the idling time $t$ by summing over the $N = 6$ cardinal states of the dual-rail Bloch sphere
\begin{equation}
    \mathcal{F}_{\mathrm{avg}}(t) = \frac{1}{N} \sum\limits_{i=1}^N \mathrm{Tr}[\rho_i(t) \rho_{i,\mathrm{ideal}}],
    \label{eq:average_state_fidelity}
\end{equation}
where $\rho_i(t)$ is the density matrix given in Eq.\,\ref{rho_no_jump} and $\rho_{i,\mathrm{ideal}}$ is the ideal density matrix. For the identity operation, the ideal density matrix is identical to the initial state. The trace in Eq.\,\ref{eq:average_state_fidelity} is 
\begin{equation}
    \begin{split}
        \mathrm{Tr}[\rho(t) &\rho_{i,\mathrm{ideal}}] = \frac{1}{1 - |v|^2 (1 - e^{- \Delta \kappa t})} \\
        &\left(|v|^4 e^{-\Delta \kappa t} + 2|u|^2|v|^2 e^{-\frac{1}{2}\Delta \kappa t- \gamma_\varphi t} + |u|^4 \right)
    \end{split}
\end{equation}
and, hence, the average state fidelity is
\begin{equation}
    \mathcal{F}_{\mathrm{avg}} = \frac{2}{6} + \frac{2}{6} \frac{\left(1 + 2 e^{- \frac{1}{2} \Delta \kappa t - \gamma_\varphi t} + e^{- \Delta \kappa t} \right)}{1 + e^{-\Delta \kappa t}}.
\end{equation}
For short timescales $\Delta \kappa t \ll 1$ and $\gamma_{\varphi} t \ll 1$, we can expand the average state fidelity
\begin{equation}
    \mathcal{F}_{\mathrm{avg}} \approx 1 - \frac{1}{3} \gamma_{\varphi} t - \frac{2}{3} \left(\frac{1}{4}\Delta \kappa t \right)^2,
\end{equation}
demonstrating that the no-jump backaction is only a second-order effect. The numerical factor in front of the second term is due to the averaging over the six cardinal states on the Bloch sphere in combination with the invariance of the logical code words to the effect of the no-jump backaction and is not listed in Table\,\ref{Tab_errors_idling_time} for clarity. 
Note that the states $\ket{01}$ and $\ket{10}$ are not affected by no-jump backaction.

\section{Derivation of the controlled joint-parity unitary}
\label{APP:Section:CJP}

The controlled-joint-parity unitary is the building block of both the joint-parity measurement used for erasure check, and the ZZ($\pi/2$) gate. This particular implementation puts the transmon ancilla in a superposition of $\ket{g}$ and $\ket{f}$ during a cavity-cavity beamsplitter interaction, allowing the transmon to effectively interact with both cavities whilst only being dispersively coupled to one. 
The joint-parity operator for two cavity modes $a$ and $b$ is $e^{i\pi(a^\dagger a + b^\dagger b)}$ and hence the desired unitary for controlled joint-parity can be written as
\begin{equation}
    U_{\mathrm{JP}} = 1 \otimes \ket{g} \bra{g} + e^{i\pi(a^\dagger a + b^\dagger b)} \otimes \ket{f} \bra{f}
    \label{eq:U_JP}
\end{equation}
Here we derive this unitary from the Hamiltonian in the Heisenberg picture. This unitary can also be derived from a geometric approach as explored in Tsunoda, Teoh et al. \cite{TakaGates2022}\\
\\
We start with a detuned beamsplitter interaction and a dispersive interaction:
\begin{equation}
    \mathcal{H} = \frac{g_{\mathrm{bs}}}{2}(a^\dagger b + a b^\dagger) + \Delta a^\dagger a + \chi_{gf} a^\dagger a \ket{f} \bra{f}
\label{eq:H0}
\end{equation}
We ignore the dynamics when the transmon is in $\ket{e}$ since these instances will be flagged as an error state indicating transmon decay. We vary $\Delta$ by detuning the beamsplitter pump(s) from the resonance condition. By setting $\Delta = -\chi_{gf} / 2$ our Hamiltonian becomes
\begin{equation}
    \mathcal{H} = \frac{g_{\mathrm{bs}}}{2}(a^\dagger b + a b^\dagger) +  \frac{\chi_{gf}}{2} a^\dagger a (\ket{f}\bra{f}-\ket{g} \bra{g}).
\label{eq:CBS}
\end{equation}
We examine the unitary 
\begin{equation}
    U_\delta(t)=e^{-iH_\delta t}
\end{equation}
generated by the related Hamiltonian 
\begin{equation}
    \mathcal{H}_\delta = \frac{g_{\mathrm{bs}}}{2}(a^\dagger b + a b^\dagger) +\delta a^\dagger a.
\end{equation}
We analyze the dynamics in the Heisenberg picture, calculating the time evolution of the cavity operators $a$ and $b$
\begin{align}
    a(t) &= U_\delta^\dagger a U_\delta\\
    b(t) &= U_\delta^\dagger b U_\delta,   
\end{align}
from which we find
\begin{align}
    &a(t) =\\
    &e^{-\frac{i\delta t}{2}}\left(\left(\cos\left(\frac{\Omega t}{2}\right)-\frac{i\delta}{\Omega}\sin{\left(\frac{\Omega t}{2}\right)}\right)a-\frac{ig_{\mathrm{bs}}}{\Omega}\sin{\left(\frac{\Omega t}{2}\right)} b\right)\\
    &b(t) =\\
    &e^{-\frac{i\delta t}{2}}\left(\left(\cos\left(\frac{\Omega t}{2}\right)+\frac{i\delta}{\Omega}\sin{\left(\frac{\Omega t}{2}\right)}\right)b-\frac{ig_{\mathrm{bs}}}{\Omega}\sin{\left(\frac{\Omega t}{2}\right)} a\right) 
\end{align}
with
\begin{equation}
    \Omega = \sqrt{g_{\mathrm{bs}}^2+\delta^2}.
\end{equation}
At time $t=\frac{2\pi}{\Omega}$, we find the cavity operators 'return' to their original value with an additional phase shift:
\begin{align}
    a\left(\frac{2\pi}{\Omega}\right) &= - e^{-\frac{i\delta \pi}{\Omega}}a\\
    b\left(\frac{2\pi}{\Omega}\right) &= - e^{-\frac{i\delta \pi}{\Omega}}b    
\end{align}
This is not a global phase shift! It corresponds to a measurable rotation of both cavity modes corresponding to the unitary
\begin{equation}
    U_\delta\left(\frac{2\pi}{\Omega}\right)=e^{i\phi(a^\dagger a +b^\dagger b)},
\end{equation}
where $\phi=\pi(1-\frac{\delta}{\Omega})$. Returning to the Hamiltonian in \ref{eq:CBS}, we can now see this generates the unitary 
\begin{equation}
    U = e^{i\phi_+(a^\dagger a +b^\dagger b)}\otimes\ket{g}\bra{g}+e^{i\phi_-(a^\dagger a +b^\dagger b)}\otimes\ket{f}\bra{f},
\end{equation}
where $\phi_\pm = \pi\left(1\mp\frac{\chi_{gf}}{\sqrt{4g_{\mathrm{bs}}^2+\chi_{gf}^2}}\right)$.\\
\\
We adjust the strength of $g_{\mathrm{bs}}$ with our pump amplitudes such that $g_{\mathrm{bs}} = \frac{\sqrt{3}}{2}|\chi_{gf}|$
\\
\\
At this operating point, $\phi_\pm = \pm \frac{\pi}{2}$ and we are implementing the same unitary as \ref{eq:U_JP} up to a deterministic $90^\circ$ cavity rotation which is easily tracked in software.\\
\\
In summary, we have shown how to realize the unitary
\begin{equation}
    U = e^{i\frac{\pi}{2}(a^\dagger a +b^\dagger b)}\otimes\ket{g}\bra{g}+e^{-i\frac{\pi}{2}(a^\dagger a +b^\dagger b)}\otimes\ket{f}\bra{f}    
\end{equation}
which is equivalent to \ref{eq:U_JP} up to unitaries of the form $e^{i\phi(a^\dagger a +b^\dagger b)}$, which are easily performed in software. The controlled joint-parity unitary is performed by enacting Hamiltonian \ref{eq:CBS} with $\Delta=-\chi_{gf} / 2$, $g_{\mathrm{bs}} = \frac{\sqrt{3}}{2}|\chi_{gf}|$ for duration $t=\frac{2\pi}{|\chi_{gf}|}$

\section{Construction of $ZZ(\theta)_{\mathrm{L}}$ from the joint-parity unitary}
\label{APP:ZZ_construction}
Here we show how to construct the $ZZ(\theta)_{\mathrm{L}}$ entangling gate from the joint-parity unitary, $U_{\mathrm{JP}}$, and rotations on our transmon ancilla. The goal is to show the unitary sequence:
\begin{equation}
    e^{-i\frac{\pi}{4}Y_{\text{gf}}}U_{\text{JP}}e^{-i\frac{\theta}{2}X_{\text{gf}}}U_{\text{JP}}e^{i\frac{\pi}{4}Y_{\text{gf}}}
\end{equation}
as presented in Fig.\,\ref{fig:two_qubit_gate} of the main text, realizes the $ZZ(\theta)$ unitary:
\begin{equation}
     \begin{pmatrix}
1 &  &  &  \\
 & e^{i\theta} &  &  \\
 &  & e^{i\theta} &  \\
 &  &  & 1 
\end{pmatrix}
\label{eq:ZZthetaU}
\end{equation}
which acts on the joint-cavity subspace $\{\ket{00},\ket{01},\ket{10},\ket{11}\}$.
When the two cavities each belong to a rail in a different dual-rail qubit, performing a physical $ZZ(\theta)$ gate on these cavities amounts to performing a logical $ZZ(\theta)_{\mathrm{L}}$ gate on the two dual-rail qubits. \\
\\
To illustrate why this construction works, we will start with the example state $(\ket{00}+\ket{01}+\ket{10}+\ket{11})\otimes\ket{g}$ and apply each unitary sequentially. A similar treatment is found in Ref.\,\cite{Gao_2019} in the construction of the eSWAP gate.
\begin{enumerate}
    \item The first $\frac{\pi}{2}$ pulse puts the transmon in a superposition in the g-f-manifold\\
    \begin{equation}
        (\ket{00}+\ket{01}+\ket{10}+\ket{11})\otimes\frac{\ket{g}+\ket{f}}{\sqrt{2}}       
    \end{equation}
    \item The first control joint-parity unitary results in the states $\ket{01}$ and $\ket{10}$ aquire a $\pi$ phase if the transmon is in $\ket{f}$. So we can write the state as\\
    \begin{equation}
    (\ket{00}+\ket{11})\otimes\frac{\ket{g}+\ket{f}}{\sqrt{2}}+(\ket{01}+\ket{10})\otimes\frac{\ket{g}-\ket{f}}{\sqrt{2}}         
    \end{equation}
    \item The rotation by $-\frac{\theta}{2}$ about the X axis is where we imprint the phase on the cavity states. If the transmon is in the $\ket{+}_\text{gf}$ state, it remains in that state but aquires phase $e^{i\frac{\theta}{2}}$. Similarly the $\ket{-}_\text{gf}$ state acquires phase $e^{-i\frac{\theta}{2}}$. After this rotation the states are\\
    \begin{equation}
    e^{-i\frac{\theta}{2}}(\ket{00}+\ket{11})\otimes\ket{+}_\text{gf}+e^{+i\frac{\theta}{2}}(\ket{01}+\ket{10})\otimes\ket{-}_\text{gf}        
    \end{equation}
    \item Now we disentangle the transmon ancilla from the cavities by applying another control-joint-parity unitary\\
    \begin{equation}
    \left(e^{-i\frac{\theta}{2}}(\ket{00}+\ket{11})+ e^{+i\frac{\theta}{2}}(\ket{01}+\ket{10})\right)\otimes\ket{+}_\text{gf}        
    \end{equation}
    \item Finally we do a $-\frac{\pi}{2}$ pulse to put the transmon ancilla back in $\ket{g}$ and then measure the ancilla to error-detect the gate
\end{enumerate}
Overall, we see the unitary we have applied to the cavities is 
\begin{equation}
     \begin{pmatrix}
e^{-i\theta/2} &  &  &  \\
 & e^{i\theta/2} &  &  \\
 &  & e^{i\theta/2} &  \\
 &  &  & e^{-i\theta/2} 
\end{pmatrix}    
\end{equation}
which is equivalent to \ref{eq:ZZthetaU} up to global phase shifts

\section{Transmon errors during joint-parity measurements and $ZZ(\theta)$ gates}
Here we show how our constructions for the joint-parity measurement and $ZZ(\theta)$ gate are robust to transmon ancilla errors during the protocol. Transmon decay generally leads to unwanted dynamics that are highly dependent on the exact time transmon decay happens, and would directly lead to Pauli errors in our dual-rail qubits. The sole reason for operating the transmon in the g-f-manifold is so we can reserve the $\ket{e}$ level for detecting transmon decay events which are then flagged as erasure.\\
\\
We are able to flag transmon decays from $\ket{f}$ to $\ket{e}$ that happen at any stage in our protocol. If the transmon decays twice or if the transmon decays to $\ket{e}$ and we misassign this as $\ket{g}$ in the measurement, then our dual-rail qubit will suffer a Pauli error, which is predicted to occur with probability $\sim10^{-4}$ per operation with state-of-the-art hardware. In both protocols, the transmon spends most of the time in an equator state such as $(\ket{g}+\ket{f})/\sqrt{2}$.  No-jump backaction will slightly polarize the transmon towards $\ket{g}$. This is already a second-order error but this can be completely counteracted by slightly over-rotating the transmon in the first $\pi/2$-pulse.
\\
\\
Transmon dephasing can be modelled by the dephasing jump operator $\sigma_z$ in the g-f-manifold. Transmon dephasing is most likely to occur during the control-joint-parity unitaries, which comprise most of the sequence length prior to transmon measurement. The Hamiltonian used to generate $U_\text{JP}$ commute with $\sigma_z$. In other words, they are error transparent to dephasing – applying the jump operator during the evolution is equivalent to applying the jump operator at the beginning or end. This property tells us that transmon dephasing during the joint-parity measurement leads to an incorrect measurement outcome and no backaction on the cavities. These measurement errors are benign. Either we unnecessarily erase a valid dual-rail qubit state or we do not flag a leakage state (such as $\ket{00}$) as erasure. The latter is a second-order error.  
\\
\\
In the $ZZ(\theta)$ construction, we apply $U_\text{JP}$ twice. If dephasing occurs in the first $U_{\text{JP}}$, it is equivalent to initializing the ancilla in $\ket{f}$ instead of $\ket{g}$ at the start of the protocol. We do the incorrect unitary $ZZ(-\theta)$ on the cavities but measure the transmon in $\ket{f}$ at the end of the protocol.  If dephasing occurs in the last $U_{\text{JP}}$, we perform the correct gate in the cavities but also measure the transmon in $\ket{f}$ and need to erase the dual-rail qubit. Finally, we note that dephasing during the short-duration transmon rotations is benign for both the joint-parity measurement and $ZZ(\theta)$ gates. In the joint-parity measurements this again results in measurement assignment errors. For the $ZZ(\theta)$ gate we find the ancilla is always mapped to $\ket{g}$ when the correct gate unitary is applied to the cavities. This is discussed further in Tsunoda, Teoh et al \cite{TakaGates2022}. We are generally not robust to cavity decay errors during the $ZZ(\theta)$ gate and instead rely on joint-parity measurements afterwards to catch this error.  
\section{Equivalence between ZZ($\pi/2$) and CZ}
\begin{figure}[t]
	\includegraphics[width=0.9\linewidth] {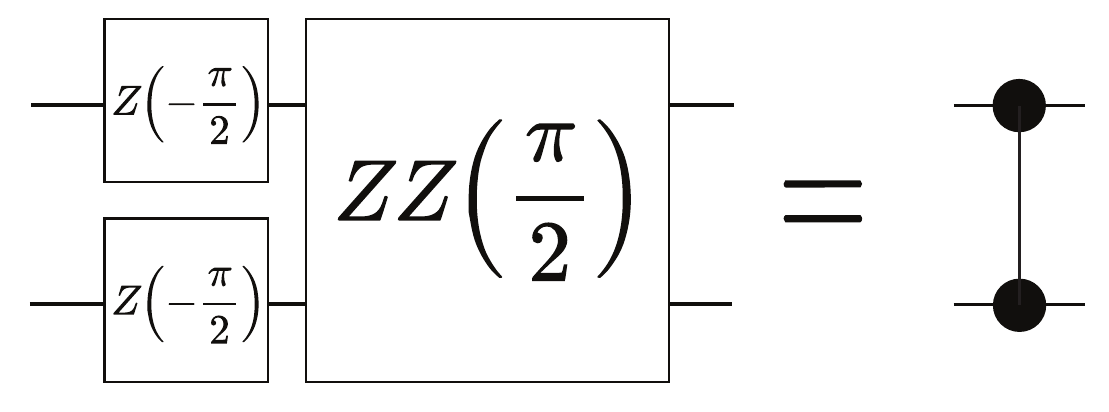}
	\caption{ 
        Local equivalence of $ZZ(\pi/2)$ and $CZ$ gates.     
}
		\label{fig:Sfig_ZZ_CNOT}
\end{figure}

 The $ZZ(\pi/2)$ gate is locally equivalent to a CZ and CNOT gate\,\cite{Schuch_2003}. With the addition of single qubit gates, we can transform $ZZ(\pi/2)$ into a CZ gate as shown in Fig.\ \ref{fig:Sfig_ZZ_CNOT}.   
\\
\\
The additional $Z(-\pi/2)$ gates can be absorbed into any adjacent single qubit gates when compiling the gate sequence. With two more Hadamard gates, we can transform the CZ gate into a CNOT gate.
\section{Two-mode SNAP gates as an alternative to the $ZZ(\theta)$ gate}
If we have another transmon ancilla dispersively coupled to two cavities in neighbouring dual-rail qubits, we can implement a logical CZ gate via two-mode SNAP. Physically, this means imparting a $\pi$ phase shift on the state $\ket{11}$. This would be an extension of the fault-tolerant SNAP protocol implemented in \cite{Reinhold2020} to the two-mode case, and requires $\chi$'s to be incommensurate such that we can resolve the joint cavity number peaks corresponding to $\ket{00},\ket{10},\ket{01},\ket{11}$ in the transmon's spectrum. This gate would have the same error detection properties as our $ZZ(\theta)$ gate.

\section{$\mathbf{ZZ(\theta)}$ gate with built-in erasure checks}
\label{APP:ZZ_with_detection}
Here we describe a modified version of the $ZZ(\theta)$ sequence which can detect the most likely cavity error: single photon loss. In this construction, the ancilla is mapped to $\ket{f}$ (and erasure is flagged) if the total number of photons in the four physical cavities is odd at the start of the gate. For instance, if one of the dual-rail qubits has previously leaked to $\ket{00}$, and the other is still in the codespace at the start of the gate, we will measure $\ket{f}$ and flag this error as erasure. 
\\
\\

\begin{figure}
    \centering
    \includegraphics[width=0.45\textwidth]{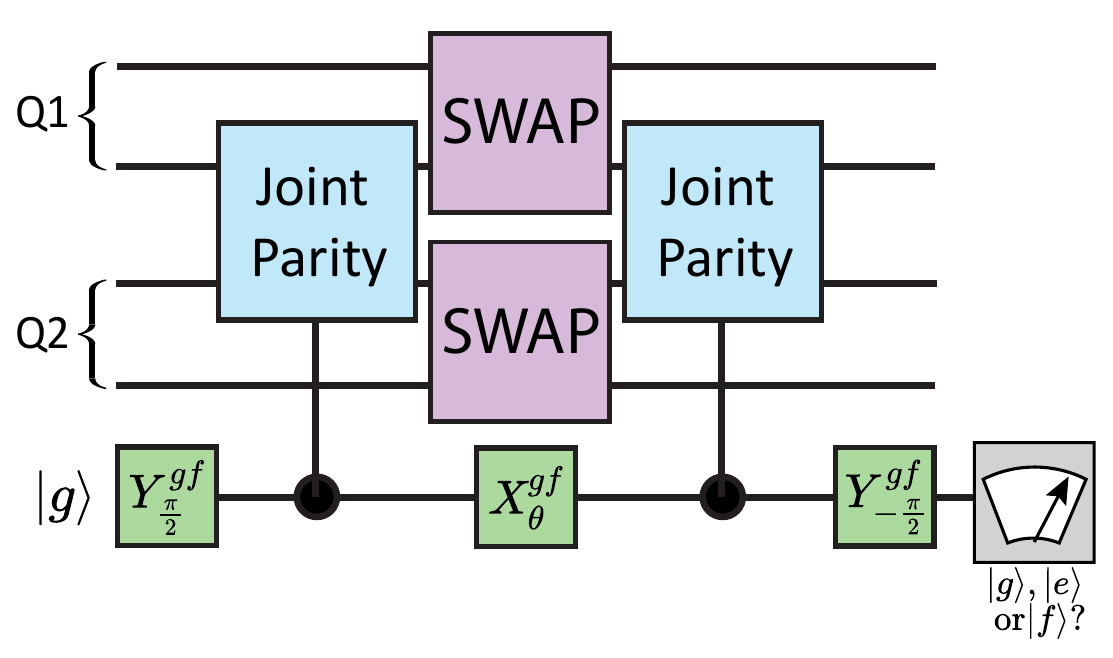}
    \caption{$ZZ(\theta)$ gate with built in error-detection for cavity leakage errors. Due to the intermediate SWAP operations, we also perform an XX gate on the dual-rail qubits. This can be easily undone with additional SWAP operations (not shown) after the gate.}
    \label{fig:ZZerrordetected}
\end{figure}
The modified ZZ sequence is shown in Fig.\ \ref{fig:ZZerrordetected} which performs the entangling gate $ZZ(\theta)$ up to an additional $ZZ_L$ operation after the gate. One of the SWAPs can be performed with just the beamsplitter interaction between two cavities but the other SWAP must be performed whilst the transmon coupled to one of the two cavities is in a superposition of $\ket{g}$ and $\ket{f}$. Despite the static dispersive interaction, it is possible to perform an ancilla state-independent SWAP as detailed in \cite{TakaGates2022}.\\
\\

We are still able to detect both transmon decay from $\ket{f}$ to $\ket{e}$ and dephasing events that happen at any point during the circuit. If photon loss occurs during the circuit, we will flag it as $\ket{f}$ half the time. If we measure $\ket{g}$ instead, we can still detect the dual-rail qubit which has leaked to $\ket{00}$ with later erasure checks. When photon loss during the sequence is undetected, the dual-rail qubit not in $\ket{00}$ may suffer a Pauli Z error. Although much rarer, if both dual-rail qubits are in $\ket{00}$ at the start of the gate we will not flag as an this error in the measurement.

\section{Physical error processes in the dual-rail hardware}
The logical errors of the dual-rail qubit are caused by physical errors in the quantum hardware, as summarized in Table\,\ref{Tab_errors_idling_time} and Table\,\ref{Tab_errors_gate} in the main text, which are a consequence of the interaction between our system and its environment. For the cavity system we consider single-photon loss and gain, and pure dephasing:
\begin{align}
    c_1 &= \sqrt{\kappa_\mathrm{a} (1 + \bar{n}_{\mathrm{th},\mathrm{a}})} a \\
    c_2 &= \sqrt{\kappa_\mathrm{b} (1 + \bar{n}_{\mathrm{th},\mathrm{b}})} b \\
    c_3 &= \sqrt{\kappa_\mathrm{a} \bar{n}_{\mathrm{th},\mathrm{a}}} a^\dagger \\
    c_4 &= \sqrt{\kappa_\mathrm{b} \bar{n}_{\mathrm{th},\mathrm{b}}} b^\dagger \\
    c_5 &= \sqrt{2 \gamma_{\varphi,\mathrm{a}}} n_\mathrm{a} \\
    c_6 &= \sqrt{2 \gamma_{\varphi,\mathrm{b}}} n_\mathrm{b}
\end{align}
Here, $\kappa_\mathrm{a}$ and $\kappa_\mathrm{b}$ are the single-photon loss rates of the cavities \textit{Alice} and \textit{Bob}, respectively. In thermal equilibrium, the decay and excitation rate for the cavities must obey detailed balance, and are therefore related by the mean number of photons $\bar{n}_{\mathrm{th},i}$, with $i \in \{\mathrm{a},\mathrm{b}\}$, which is typically equivalent to a few percent first excited state population. As a consequence, the excitation rates are roughly $10^{-2}$ lower than the decay rates. The pure dephasing rates are $\gamma_{\varphi,\mathrm{a}}$ and $\gamma_{\varphi,\mathrm{b}}$. \\
\\
For the transmon ancilla operated in the g-f-manifold, we assign transition rates between the three energy levels \cite{Morvan_2021}
\begin{align}
    c_7 &= \sqrt{\Gamma_\downarrow^{ge}} \ket{g}\bra{e} \\
    c_8 &= \sqrt{\Gamma_\uparrow^{ge}} \ket{e}\bra{g} \\
    c_9 &= \sqrt{\Gamma_\downarrow^{ef}} \ket{e}\bra{f} \\
    c_{10} &= \sqrt{\Gamma_\uparrow^{ef}} \ket{f}\bra{e} \\
    c_{11} &= \sqrt{2 \Gamma_{\varphi}^{ff}} \ket{f}\bra{f}.
\end{align}
While the matrix transition elements between neighboring states $i$ and $i+1$ in a transmon are expected to increase similar to the harmonic oscillator \cite{Koch_2007} with increasing level index for many decay mechanisms \cite{Peterer2015}, the anharmonicity of the transmon results in a different noise spectral density associated with each transition. As a consequence, the decay rate in the g-f-manifold is not necessarily higher compared to the g-e-manifold. The collapse operator introducing pure dephasing is defined such that the dephasing in the g-f-manifold is $\Gamma_{\varphi}^{gf} = \Gamma_{\varphi}^{ff}$.
\\
\\
In this analysis, we do not consider heating to the $\ket{h}$ state of the transmon, described by the collapse operator $\sqrt{\Gamma_\uparrow^{fh}}\ket{h}\bra{f}$ although its contribution is expected to be small. This is because the heating rate $\Gamma_\uparrow^{fh}$ is expected to be comparable to $\Gamma_\uparrow^{ge}$ at the $10^{-4}$ level and we are only susceptible to this error during two-qubit gates and leakage detection via joint-parity measurements. Furthermore, it is also possible to detect this error when measuring the transmon and treat it as an erasure by using four-state single-shot readout which can distinguish $g,e,f$ and $h$ as was shown in \cite{SalElder2020}.
\section{Relation to Linear Optics Quantum computing}
Many of the operations presented are the same as those required for linear optics quantum computing, which often uses the dual-rail code as well but for propagating photon modes. The state preparation, readout and beamsplitter operations we have described are by themselves sufficient to implement universal quantum computation via the KLM protocol \cite{KLM2000}. Moreover, these operations are the most robust to ancilla transmon errors. However, the KLM protocol and related proposals for optical cluster states \cite{FusionBased2021} are all measurement-based. They incur substantial overheads in the number of physical modes required and suffer from finite success probabilities in the preparation of entangled states. 
By developing a deterministic two-qubit entangling gate, we are able to do gate-based computing instead via the $ZZ(\theta)$ gate. This comes at the cost of first-order transmon errors setting our erasure rate instead of cavity photon loss, and second-order transmon errors setting the Pauli error rate.

\section{Operating the transmon in the g-f-manifold}
The erasure check and $ZZ(\theta)$ gate both use transmon ancilla rotations in the g-f-manifold in order to reserve the $\ket{e}$ state solely for detecting transmon decay during the gate.\\
\\
There are two main ways to achieve this. With pulses on resonance with the transmon transition frequencies $\omega_{ge}$ and $\omega_{ef}$, we can perform arbitrary rotations in the g-f-manifold by temporarily populating the $\ket{e}$ state \cite{WallraffTransmonQutrit2010}. Arbitrary rotations can be performed at speeds comparable to single qubit gates in the g-e-manifold of the transmon. However, we are not robust to all the first-order transmon errors that can occur during these rotations, and this limits our Pauli error to the fidelity of single qubit transmon gates (which is still $\sim10^{-4}$).\\
\\
We can circumvent this limit if we wish, by only virtually populating the $\ket{e}$ state during our $gf$ rotations. This can be done by applying two pulses at frequencies $\omega_1,\omega_2$ such that $\omega_1+\omega_2=\omega_{ge}+\omega_{ef}$. By satisfying this condition and setting $\omega_1=\omega_{ge}-\Delta_{R}$, we drive a virtual Raman transition between $\ket{g}$ and $\ket{f}$, suppressing the population of the $\ket{e}$ state by $\left(\frac{\Delta_R}{\Omega_R}\right)^2$ for the effective transition drive strength $\Omega_R$. Any leakage to the $\ket{e}$ state in this process is now also a detectable error, and first-order transmon errors during the rotation can no longer lead to an undetected Pauli error. We can also achieve the same control with a single drive frequency by setting $\omega_1=\omega_2=\frac{\omega_{ge}+\omega_{ef}}{2}$. Leakage to $\ket{e}$ can be further suppressed by using DRAG techniques for pulse shaping \cite{Motzoi_2009,Gambetta_DRAG_2011}. $\pi$-pulses in the g-f-manifold can be done in $\sim 50$ ns with this approach, with $<0.5\%$ leakage to $\ket{e}$.

\section{Sources of cavity dephasing}
The intrinsic dephasing of microwave cavity resonators is thought to be extremely low. Cavity $T_\phi$ dephasing times up to $20\mathrm{ms}$ \cite{VladGKP2022} have been reported when we mitigate the cavity dephasing caused by transmon ancilla heating events. In the circuit-QED platform, the bare cavity frequency is set entirely by the geometry of the conductors. As a result, we expect extremely stable cavity frequencies for coaxial and planar geometries, even in the presence of external mechanical vibrations.

\section{Unwanted non-linearities inherited from the transmon ancilla} 
When we operate a qubit-cavity system in the dispersive regime, we often want to engineer only a pure dispersive coupling of the form 
\begin{equation}
    \mathcal{H}_{\text{disp}}^{(1)}=\chi_{ge}a^\dagger a \ket{e}\bra{e}+\chi_{gf}a^\dagger a \ket{f}\bra{f}
\end{equation}
This is only an approximation of our full system Hamilotonian. Higher order non-linearities such as cavity self Kerr and $\chi'$ are also necessarily present in our system and can be modelled with additional Hamiltonian terms:
\begin{equation}
    \mathcal{H}_{\text{disp}}^{(2)}=\frac{K}{2}a^\dagger a^\dagger a  a+\chi'_{ge}a^\dagger a^\dagger  a  a\ket{e}\bra{e}+\chi'_{gf}a^\dagger a^\dagger  a  a\ket{f}\bra{f}
\end{equation}
$\chi'$ can be thought of as a transmon-dependent cavity Kerr. Since $\chi '\sim K \sim\chi\left(\frac{g_0}{\Delta}\right)^2$
These typically cause unwanted coherent dynamics over slower timescales than our gate operations but nonetheless have been shown to degrade the performance of Bosonically encoded qubits \cite{Campagne-Ibarcq2020,Ofek2016}.
\\
\\
We emphasize that the Dual-Rail encoding is highly resilient to these effects, for all operations. Within a single dual-rail qubit, higher order non-linearities have no effect on the dyniamics since we are restricted to the single-photon subspace. When we perform our $ZZ(\theta)$ gate on two cavities (one from each dual-rail qubits) we are slightly sensitive to $K$ and $\chi'$.
\\
\\
This is because, $1/4$ of the time, the cavities will be in the state $\ket{11}$ and during the $ZZ(\theta)$ gate we will deterministically evolve out of the dual-rail subspace towards the $\ket{20}+\ket{02}$ state before returning back to the initial $\ket{11}$ state with the desired phase acquired (of $2\pi$). $K$ and $\chi'$ will have two very small effects. Firstly, the phase acquired will not be exactly $2\pi$ but this deviation can in principle be measured and fully compensated with single-qubit gates afterwards. Secondly, there is a small probability ($\sim (KT_{\text{gate}})^2\approx\left(\frac{K}{\chi}\right)^2\sim 10^{-6}$) that we end up in the state $\ket{20}+\ket{02}$. Once again, this error is detectable via subsequent joint-parity measurements on each dual-rail qubit.  
\\
\\
With this inherent robustness to higher order Hamiltonian terms, the optimum $\chi$ to engineer for dual-rail qubits is likely higher than the typical $1 \text{ MHz}$ used for higher photon number bosonic codes \cite{Ofek2016,Rosenblum2018, Reinhold2020}. $\chi$ could be as high as $\sim 10\text{ MHz}$ making it possible to implement the $ZZ(\theta)$ gate $\sim100 \text{ ns}$ , comparable to two-qubit gates in transmons. Engineering a larger $\chi$ in hardware is easily achieved by increasing the capacitive coupling between the transmon and cavity mode \cite{HeeresSNAP2015}.\\
\\
Note as we increase $\chi$, the cavity mode will experience an enhanced photon loss due to the inverse Purcell effect, and will inherit intrinsic dephasing from the transmon which scales as $\sim \left(\frac{g_0}{\Delta}\right)^2\Gamma_\phi$. Where $g_0$ is the vacuum Rabi coupling between the transmon and the cavity mode and $\Delta$ is the detuning.
Decreasing our gate time is desirable. Not only does our erasure rate decrease linearly but we expect a quadratic reduction in Pauli errors from double transmon errors. 
\\
\\
\section{Unwanted non-linearities inherited from the beamsplitter coupler}
The beamsplitter coupler will likely contain at least one Josephson junction and hence both the cavities bridged by the coupler will likely inherit self-Kerr and cross-Kerr terms, the extent of which depends on the exact hardware implementation used for the coupler. We have shown in the previous section that cavity self-Kerr is benign for the Dual-rail code. However, a static cross-Kerr coupling of the form
\begin{equation}
    \mathcal{H}_{\text{Cross-Kerr}}=\chi_{ab}a^\dagger a b^\dagger b,
\end{equation}
for cavity modes $a$ and $b$. This interaction only causes unwanted effects when we bridge cavities between neighbouring dual-rail qubits (not the two cavities within a single dual-rail qubit) because a single photon may be present in \emph{both} cavities. This is very similar to the stray ZZ interactions found in transmon qubit architectures and we can cancel this interaction during idle times with dynamical decoupling. Alternatively,  couplers with tunable cross-Kerr have already been realized \cite{StijnBS2022} and we may completely null the cross Kerr instead to mitigate this problem.  
\\
\\
The effect of $\chi_{ab}$ during a $ZZ(\theta)$ gate is also rather benign. Similarly to the effects from $\chi'$ and $K$, there is a small probability of leaking to ${\ket{02},\ket{20}}$ manifold and slightly altered phase accumulation on the $\ket{11}$ state. 

\section{Proposed state preparation via sideband drives}
Sideband drives to swap excitations from the transmon ancilla to the cavity can be used to quickly load a photon into a microwave cavity. The preparation itself may be susceptible to transmon errors, so long as this only introduces additional erasure errors at the $<10^{-2}$ level and Pauli errors at the $<10^{-4}$ level. Preparation can be further vetted with subsequent parity measurements on the cavity. \\
\\
The sideband interaction can be parametrically activated by applying off-resonant drive tone(s) to the transmon to activate a four-wave-mixing process. We can apply two tones at frequency $\omega_x$ and $\omega_y$. For a cavity mode at frequency $\omega_a$ and a transmon with transition frequencies $\omega_{ge}$ and $\omega_{gf}$, when $|\omega_x-\omega_y|=|\omega_{a}-\omega_{ge}|$, we engineer the Hamiltonian
\begin{equation}
    \mathcal{H} = \frac{\Omega}{2} \left(a^\dagger \ket{g}\bra{e} + a \ket{e}\bra{g}\right)
\end{equation}
Which effectively puts the cavity mode on resonance with the transmon in the g-e-manifold. By initially preparing the transmon in $\ket{e}$, we can fully swap this excitation to a single photon in the cavity by applying the drives for time $t = \frac{\pi}{\Omega}$. There is a chance the cavity remains in vacuum due to transmon errors, but this is a detectable error.\\
\\
Alternatively, we may apply a single drive at frequency to turn on the interaction 
\begin{equation}
    \mathcal{H} = \frac{\Omega}{2} \left(a^\dagger \ket{g}\bra{f} + a \ket{f}\bra{g}\right)
    \label{eq:gf_sideband}
\end{equation}
Which requires only a single tone at frequency $|\omega_{gf}-\omega_c|$ and the ability  to prepare the transmon in $\ket{f}$. This enhances $\Omega$ by a factor of $\sqrt{2}$ and uses one less pump tone compared to the first approach. This was realized in \cite{Rosenblum2018CNOT} with $\Omega \approx \text{10 MHz}$\\
\\
A step-by-step protocol to load a photon could be:
\begin{enumerate}
\item Prepare the transmon and cavity in their ground states via measurements and feedback \cite{DelftTransmonReset2012}
\item Flip the transmon to $\ket{f}$ with a $g$-$e$ $\pi$-pulse followed by an $e$-$f$ $\pi$-pulse. (Alternatively a direct $g$-$f$ $\pi$-pulse via a two-photon drive would also suffice.)
\item Turn on the sideband interaction in \ref{eq:gf_sideband} for time $t=\frac{\pi}{\Omega}$
\end{enumerate}

\end{document}